\documentclass{mn2e}
\def\eg{\hbox{e.g. }}
\def\etal{\hbox{et al. }}
\def\spose#1{\hbox to 0pt{#1\hss}}
\def\approxgt{\mathrel{\spose{\lower 3pt\hbox{$\sim$}}\raise 2.0pt\hbox{$>$}}}
\def\approxlt{\mathrel{\spose{\lower 3pt\hbox{$\sim$}}\raise 2.0pt\hbox{$<$}}}

\usepackage{graphics,graphicx}

\newcommand{\ion}[2]{#1\,{\sc{#2}}}

\title[Core-collapse supernova enrichment in M\,87]
{Core-collapse supernova enrichment in the core of the Virgo Cluster}
\author[E. T. Million \etal]
{\parbox[]{6.in} {E. T. Million$^{1,2,3}$, N. Werner$^2$, A. Simionescu$^2$, \& S. W. Allen$^{2,3}$\\
\footnotesize
$^1$University of Alabama, 206 Gallalee Hall, Box 870324, Tuscaloosa, AL, 35487, USA\\
$^2$Kavli Institute for Particle Astrophysics and Cosmology, Stanford University, 382 Via Pueblo Mall, Stanford, CA 94305-4060, USA\\
$^3$SLAC National Accelerator Laboratory, 2575 Sand Hill Road, Menlo Park, CA 94025, USA\\
}}
\begin{document}
\renewcommand{\thefootnote}{\arabic{\footnote}}
\maketitle
\begin{abstract}
Using a deep (574~ks) {\it Chandra} observation of M\,87, the dominant galaxy of the nearby Virgo Cluster, we present the best measurements to date of the radial distribution of metals in the central intracluster medium (ICM). Our measurements, made in 36 independent annuli with $\sim$250,000 counts each, extend out to a radius $r\sim40$~kpc and show that the abundance profiles of Fe, Si, S, Ar, Ca, Ne, Mg, and Ni are all centrally peaked. 
Interestingly, the abundance profiles of Si and S - which are measured robustly and to high precision - are even more centrally peaked than Fe, while the Si/S ratio is relatively flat. These measurements challenge the standard picture of chemical enrichment in galaxy clusters, wherein type Ia supernovae (SN~Ia) from an evolved stellar population are thought to dominate the central enrichment. The observed abundance patterns are most likely due to one or more of the following processes: continuing enrichment by winds of a stellar population pre-enriched by SN$_{\rm CC}$ products; intermittent formation of massive stars in the central cooling core; early enrichment of the low entropy gas. 
We also discuss other processes that might have contributed to the observed radial profiles, such as a stellar initial mass function that changes with radius; changes in the pre-enrichment of core-collapse supernova progenitors; and a diversity in the elemental yields of SN~Ia. 
Although systematic uncertainties prevent us from measuring the O abundance robustly, indications are that it is about 2 times lower than predicted by the enrichment models. 

\end{abstract}

\begin{keywords}
galaxies: abundances -- galaxies: individual: M\,87 -- 
X-rays: galaxies: clusters -- galaxies: clusters: intracluster medium
\end{keywords}

\section{Introduction}

The discovery of Fe-K line emission in galaxy clusters (Mitchell \etal 1976; 
Serlemitsos \etal 1977) showed that their X-ray emitting intracluster medium (ICM) contains
significant amounts of processed elements created by supernovae. 
Different types of supernovae synthesize different ratios of elements.
In particular, type Ia supernovae (${\rm SN\;Ia}$) produce large amounts of Fe and Ni.
Meanwhile, core-collapse supernovae (${\rm SN_{\rm CC}}$) 
produce lighter elements such as O, Ne and Mg. 
Si-group elements (Si, S, Ar, and Ca) are produced by both supernova types. 
 By measuring specific elemental abundances 
within the ICM, we can separate the relative contributions to the chemical enrichment
by different types of supernovae (\eg Werner \etal 2008;
B\"ohringer \& Werner 2010 and references therein).
Such measurements place important constraints on models of supernova
explosions (\eg Dupke \& White 2000; de Plaa \etal 2007; Simionescu \etal 2009).

Previous results for clusters and groups pointed to a centrally peaked Fe abundance distribution,
coupled with a relatively flat O abundance profile, in cool core systems.
Based on such measurements, it was proposed that the relative contribution of ${\rm SN\;Ia}$ to the enrichment of clusters increases
towards their central regions (Finoguenov \etal 2000; 
B\"ohringer \etal 2001; Tamura \etal 2001; Finoguenov \etal 2002; 
Matsushita \etal 2003). The central Fe abundance peaks in cool core clusters were expected to
form primarily by ${\rm SN\;Ia}$ with long delay times, as well as stellar mass loss in the cD galaxies
(B\"ohringer \etal 2004; De Grandi \etal 2004).
${\rm SN_{CC}}$ products such as O, Ne, and Mg were, on the other hand, produced early in the formation history
of clusters at $z\sim2$--3 and were thought to be well mixed throughout the ICM. Because ${\rm SN_{CC}}$ also produce large amounts of Si and S, this scenario predicts shallower central gradients for the distributions of Si and S than for Fe, resulting in Si/Fe and S/Fe ratios that increase with radius. 

The predicted Si/Fe gradient has, however, not been seen and the observed radial profiles of the Si/Fe ratios are in most clusters consistent with being flat, indicating that Si is just as peaked as Fe (\eg
B\"ohringer \etal 2001; Finoguenov \etal 2002; Tamura \etal 2004; 
Sanders \etal 2004; Durret \etal 2005; de Plaa \etal 2006; Werner \etal 2006a; Sato
\etal 2007, 2008; Matsushita \etal 2007).
Moreover, contrary to previous claims, Simionescu \etal (2009) show that in Hydra A and in a sample of other nearby clusters of galaxies the O abundance is also centrally 
peaked.  
The observed centrally peaked distribution of ${\rm SN_{CC}}$ products suggests that 
the metallicity gradients form early in the history of clusters and persist for a long time, or that the
enrichment by the winds of the evolved stellar population is significantly more efficient than previously thought (Simionescu \etal 2009).
Centrally peaked ${\rm SN_{CC}}$ products are also observed in the core of the Centaurus Cluster
and have been interpreted as being either due to continuous or intermittent star formation over the past $\sim$8 Gyr, 
or due to the early enrichment during the formation of the central galaxy (see Sanders \& Fabian 2006).

This is the third in a series of papers analysing a deep (574 ks) {\it 
Chandra} observation of M\,87, the dominant galaxy of the Virgo Cluster. The first two papers 
(Million \etal 2010, hereafter Paper I; Werner \etal 2010, hereafter Paper II)
focus on the effect of the AGN on the surrounding hot plasma.
This paper focuses on the 
history of chemical enrichment of the central regions of the Virgo Cluster.
Many previous studies of this subject utilize
{\it XMM-Newton} or {\it Suzaku} because of their better spectral resolution.
However, the far superior spatial resolution of {\it Chandra}, which
allows us to better separate out obvious complications due to
substructure (as well as the sheer number of photons already collected
with {\it Chandra}) allow us to probe the detailed distribution of metals 
with greatly improved accuracy over previous work.

The structure of this paper is as follows. Sect. 2 describes the data
reduction and spatially resolved spectroscopy. Sect. 3 presents
our detailed abundance and abundance ratio profiles. 
Sect. 4 describes the implications of our measured radial abundance profiles on 
the history of chemical enrichment through supernovae. Sect. 5 summarizes the results and the conclusions.

Throughout this paper, we assume that the cluster lies at a distance of 
16.1 Mpc (Tonry \etal 2001), for which the linear scale is 0.078 kpc per arcsec.

\section{Data reduction and analysis}

\begin{figure}
\scalebox{0.43}{\includegraphics[angle=270]{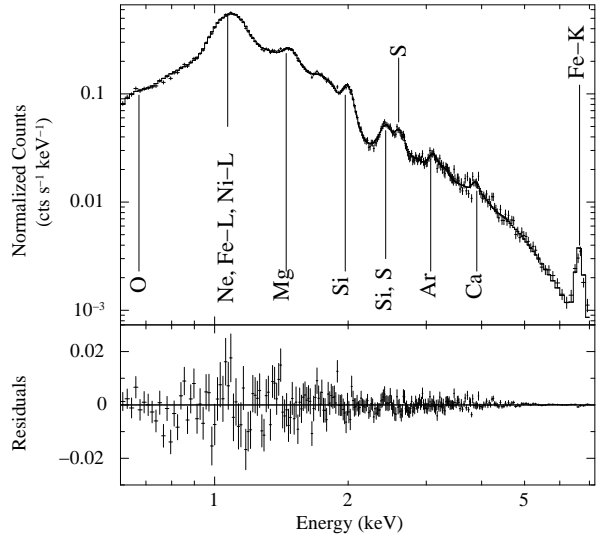}}
\caption{Example spectrum and residuals from the radial analysis.
The best-fit {\small APEC} model is overplotted
as a solid line and is a good fit to the data.
The line contributions of O, Ne, Mg, Si, S, Ar, Ca, Fe, and Ni
are labeled on the plot 
and are easily observable in spectra with $\sim$250,000 net counts. 
The lower panel shows the residuals to the fit. }
\label{fig:spectra}
\end{figure}

\subsection{Data reduction}

A detailed description of the data reduction and analysis is
described in Paper I. 
The {\it Chandra} observations were carried out using the Advanced CCD
Imaging Spectrometer (ACIS) between July 2002 and November 2005.  
The net exposure time after cleaning is 574~ks.
Here, we use an updated version of 
the {\it CIAO} (version 4.3) software package, including the 
appropriate gain maps and updated calibration products ({\it CALDB} version 4.4.1).

\subsection{Spectral analysis}
\label{section:annulus}

In order to minimize systematic uncertainties associated with the 
multi-temperature structure of the X-ray emitting gas, 
we conservatively exclude the X-ray bright arms and the innermost multiphase core
(more details can be found in Paper I). 
As discussed in Paper I, beyond the innermost core and the X-ray bright arms, the gas can be
well described by an isothermal model at each radius.
We use partial annuli that vary in width from
10 to 30 arcsec, with $\sim$250,000 net counts in each of a total of 36 regions. 

Background spectra were determined using the blank-sky fields
available from the Chandra X-ray Center (see Paper I for details). 
Due to the high X-ray brightness of the target, uncertainties in the
background modeling have little impact on the determination of the quantities
presented here.

The spectra have been analyzed using {\small XSPEC} (version 12.5; Arnaud
1996). Each annulus is fit with a photoelectrically absorbed (Balucinska-Church \& McCammon
1992) single temperature APEC thermal plasma model (Smith \etal 2001; AtomDB v2.0.1 was used). 
We fix the Galactic absorption to $1.93\times10^{20}$~cm$^{-2}$
(determined from the {\it Leiden/Argentine/Bonn}
radio \ion{H}{i} survey; Kalberla \etal 2005). In order to investigate modeling uncertainties in the determination of chemical abundances, we also repeated the spectral fits using the MEKAL thermal plasma model (Kaastra \& Mewe 1993; Liedahl \etal 1995).
All spectral fits were carried out in the $0.6-7.0$ keV energy band. The extended C-statistic available in {\small XSPEC},
which allows for background subtraction, was used for all spectral fitting.

The temperature, the normalization, and the abundances of O, Ne, Mg, Si, S, Ar, Ca, Fe, and Ni are
free parameters for every annulus. Fig. \ref{fig:spectra} shows an example
spectrum containing $\sim$250,000 net counts with the metal line emission labeled.  
The spectra allow us to determine the Fe abundance 
to within $\leq3$ per cent, the abundances of Si and S to within $\leq5$
per cent, the abundances of Ne, Mg, and Ni to within $\leq10$ per cent,
and the abundances of Ar and Ca to within $\leq20$ per cent 
statistical precision. 
The errors on the abundance profiles were determined from a Markov Chain Monte Carlo (MCMC) analysis.
Measurement errors were determined from the 68 per cent confidence posterior distribution
of the MCMC analysis. Chain lengths were at least $10^4$ samples after correcting for burn-in.

Abundances in the paper are given
with respect to the `proto-Solar' values of Lodders (2003).

\subsubsection{The AtomDB v2.0.1 atomic database}

Our modelling makes use of the recently updated AtomDB v2.0.1 atomic database used by the {\small APEC} thermal plasma model implemented in {\small XSPEC}. 
This represents a major update from the previous AtomDB v1.3.2 with nearly all atomic data replaced.\footnote{www.atomdb.org} 
The update of the atomic database affects most significantly the Fe abundance, which is on average lower by $\sim20$ per cent in v2.0.1 compared to v1.3.2. 
This change has a slight dependence upon the temperature of the plasma. The abundances of Si, S, Ar, Ca, Ne, Mg, and Ni are smaller by less than $\sim$10 per cent in the updated version. The abundance ratios with respect to Fe are significantly higher as a result. We note that our main conclusions based on the centrally peaked abundances and abundance ratios are not sensitive to our choice of plasma code.

\section{Metallicity of the X-ray gas}
\label{section:profiles}

\begin{figure*}
\scalebox{0.32}{\includegraphics[angle=270]{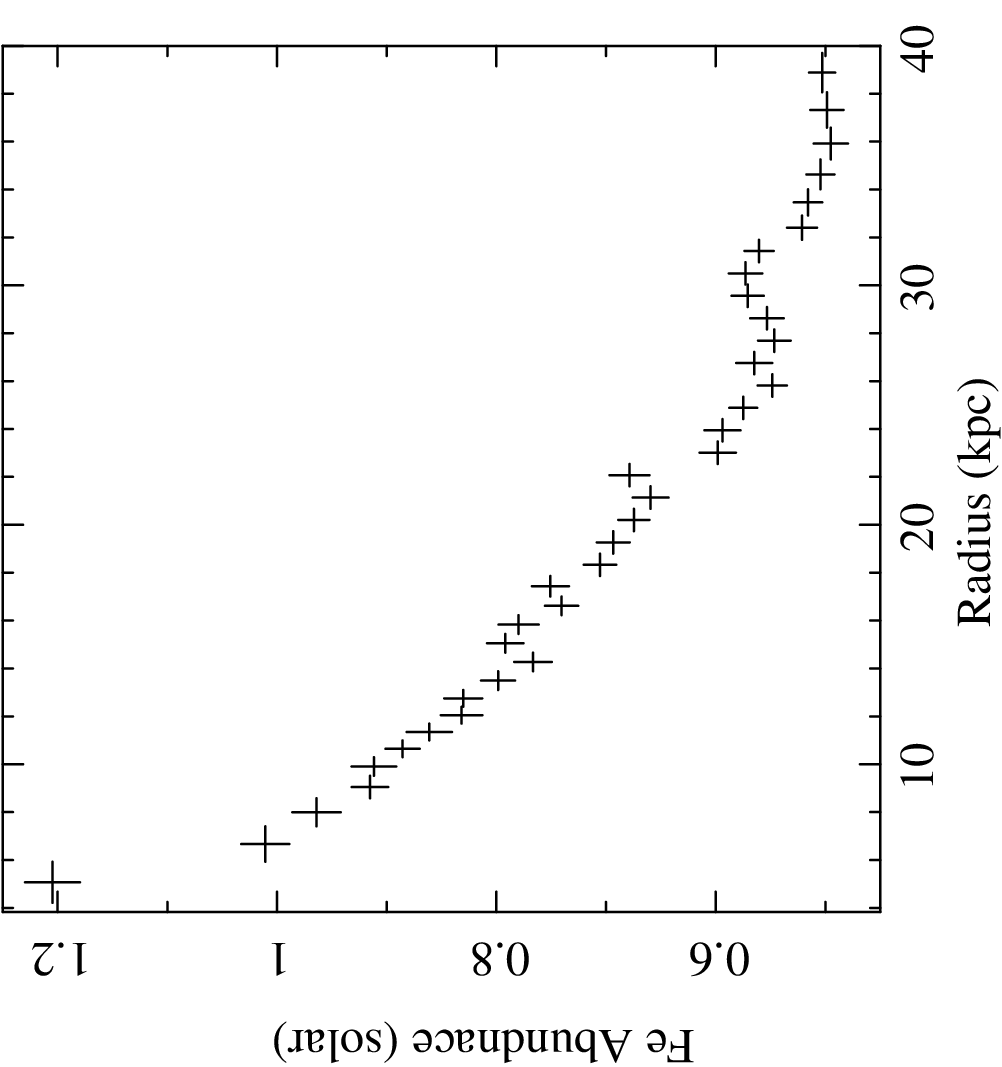}}
\scalebox{0.32}{\includegraphics[angle=270]{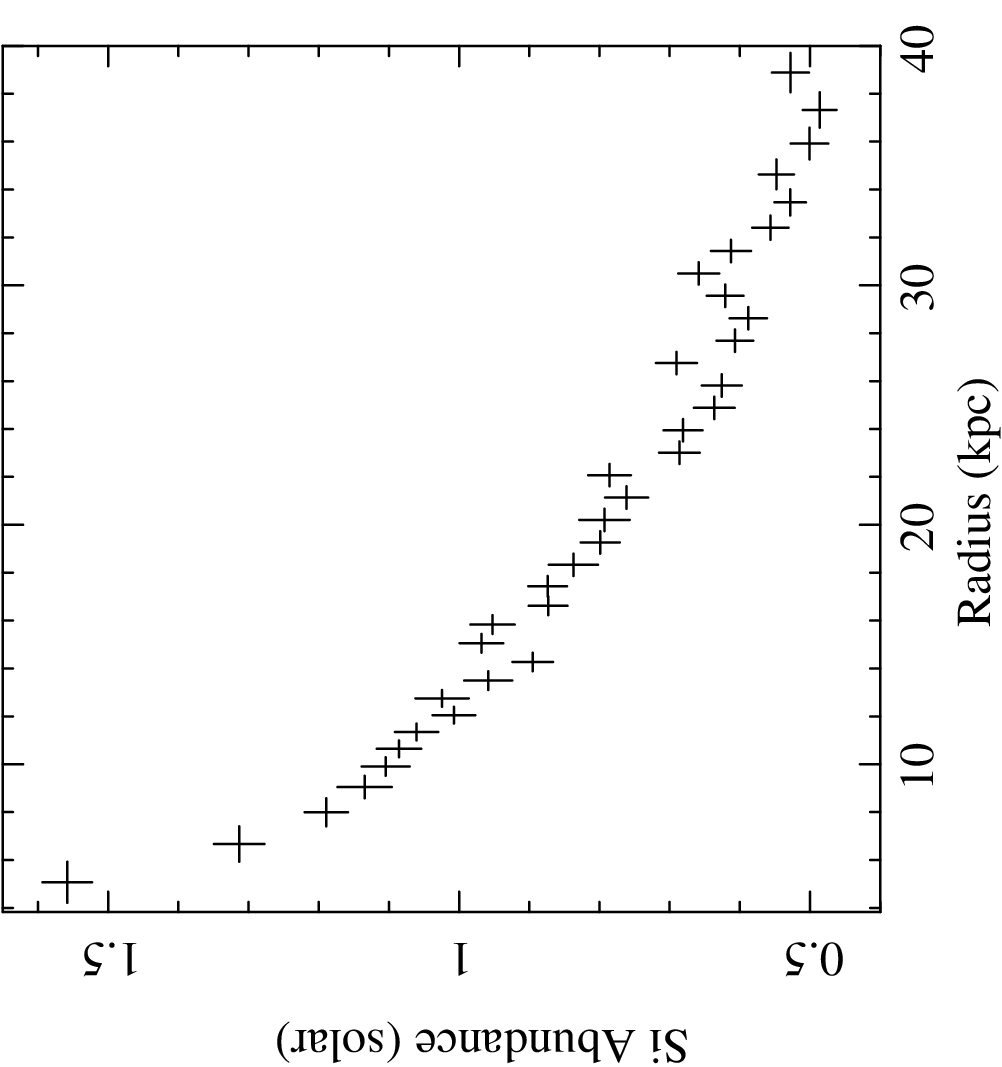}}
\scalebox{0.32}{\includegraphics[angle=270]{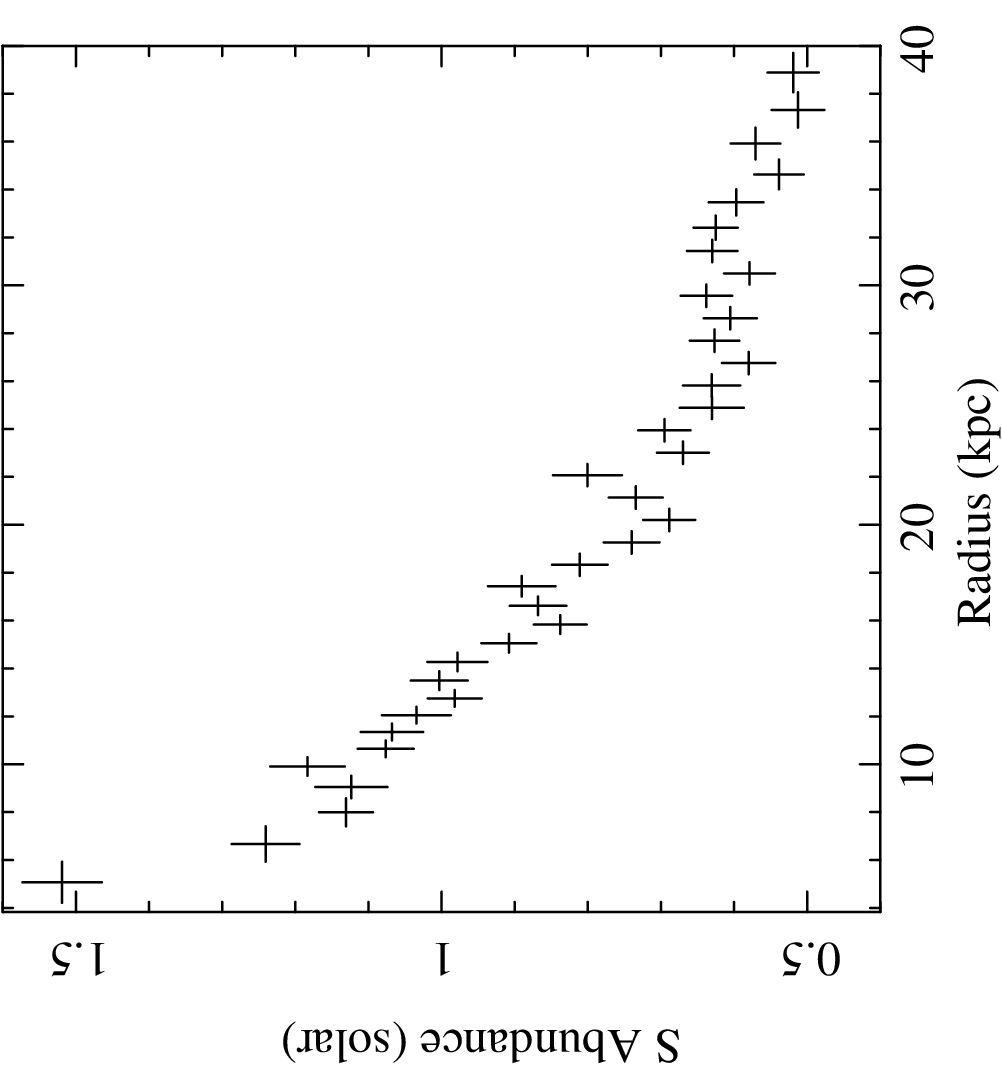}}
\scalebox{0.32}{\includegraphics[angle=270]{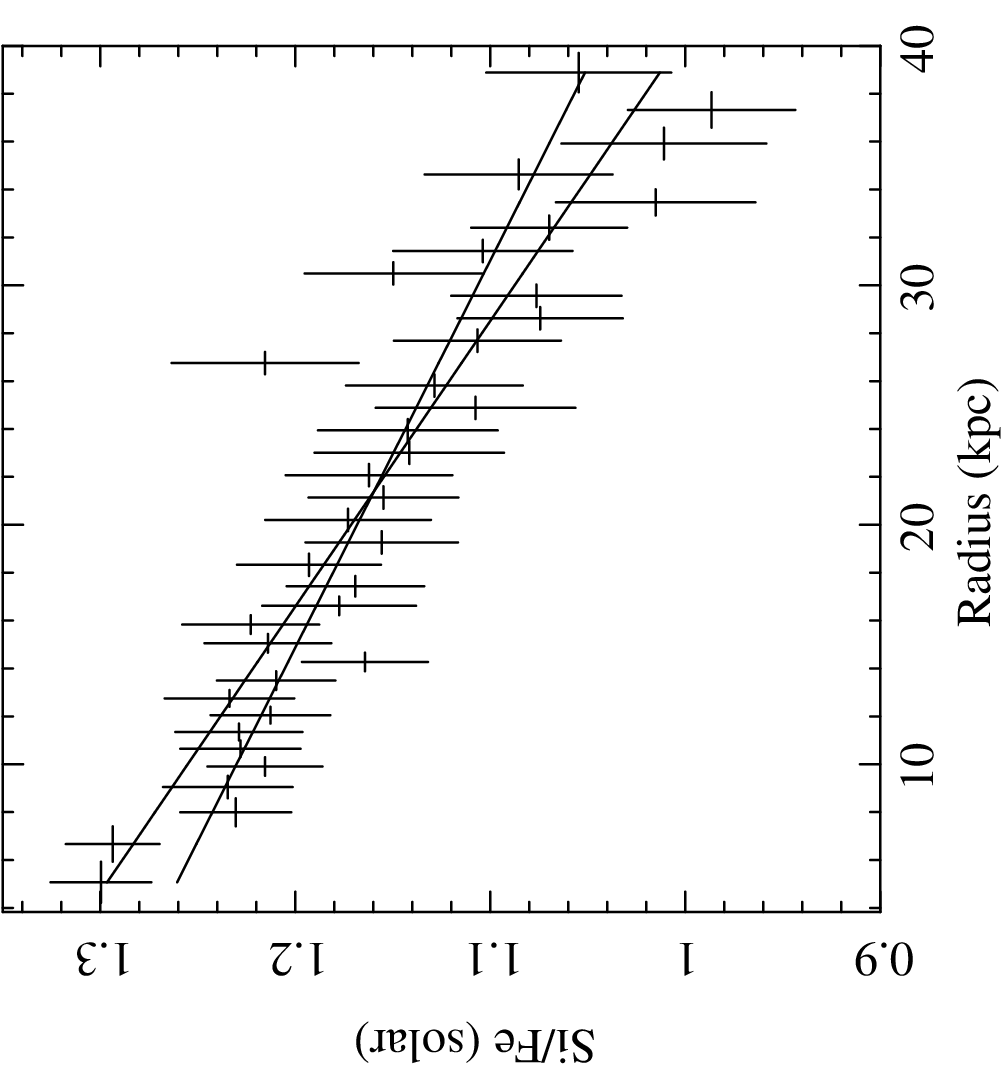}}
\scalebox{0.32}{\includegraphics[angle=270]{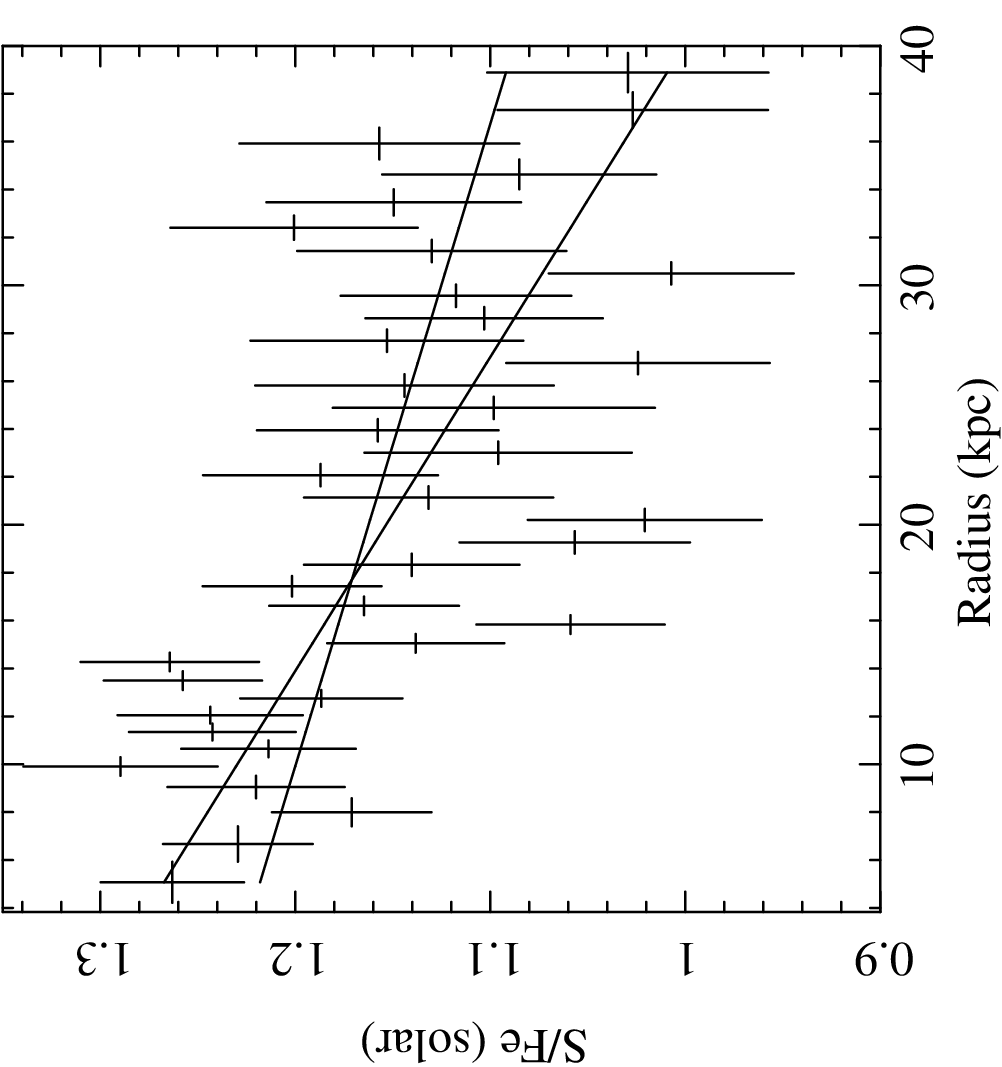}}
\scalebox{0.32}{\includegraphics[angle=270]{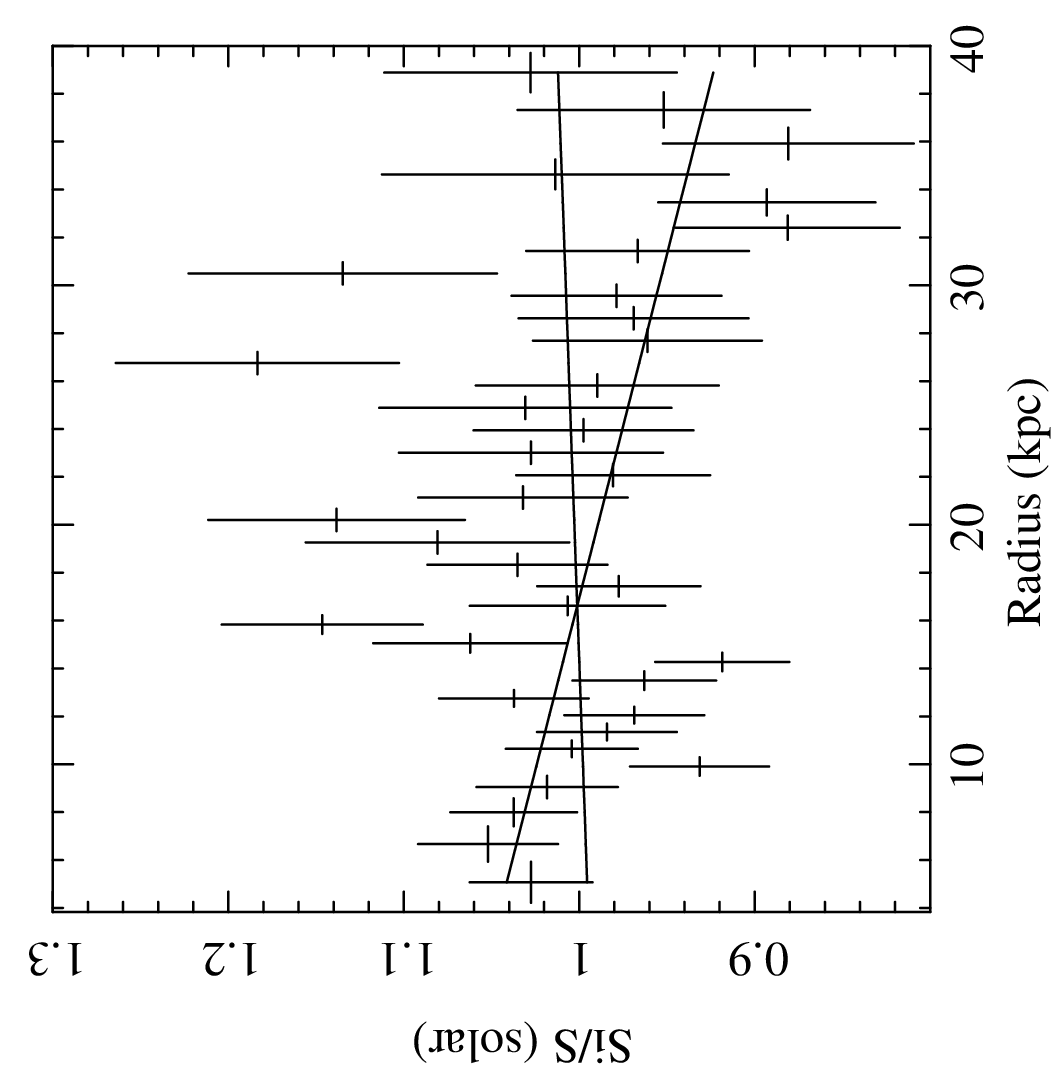}}
\caption{Top row: Abundance profiles of Fe (left; see also Paper I), 
Si (middle), and S (right). Bottom row: Abundance ratio profiles of Si/Fe (left),
S/Fe (middle), and Si/S (right). Fe, Si, and S have the best determined abundances.
The 95 per cent lower and upper limits of the slopes of the radial distributions of the abundance ratios are overplotted. 
}
\label{fig:FeSiS}
\end{figure*}

\begin{figure*}
\hspace{1.5cm}\scalebox{0.32}{\includegraphics[angle=270]{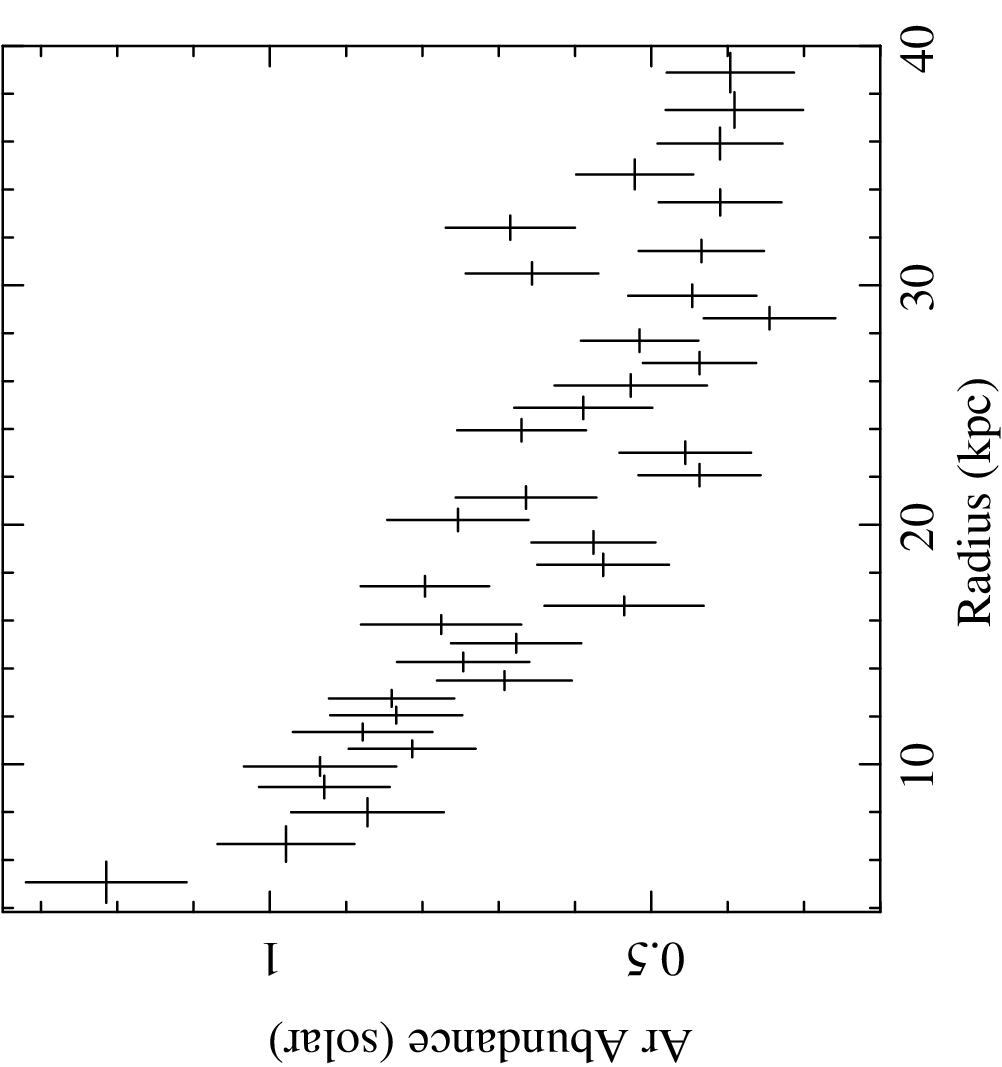}}
\hspace{1.7cm}\scalebox{0.32}{\includegraphics[angle=270]{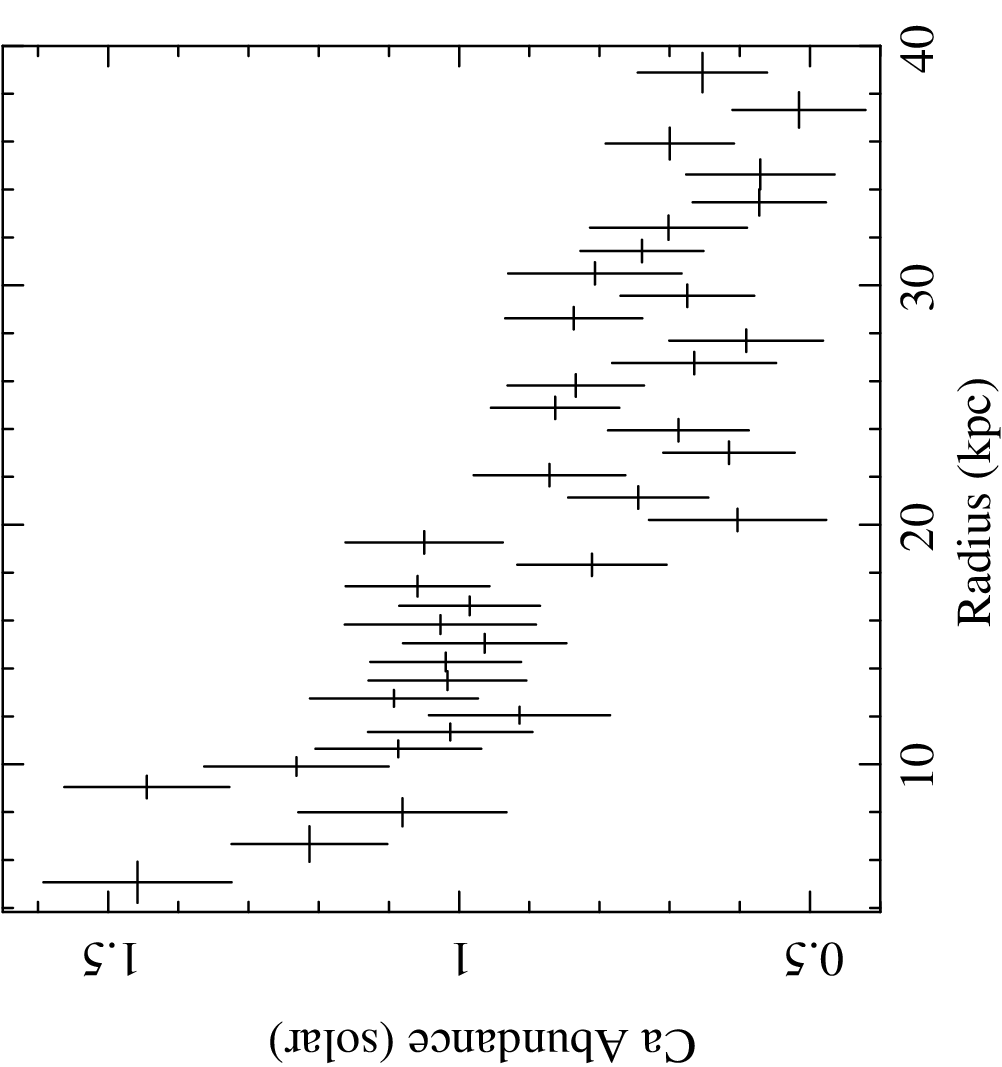}}
\scalebox{0.32}{\includegraphics[angle=270]{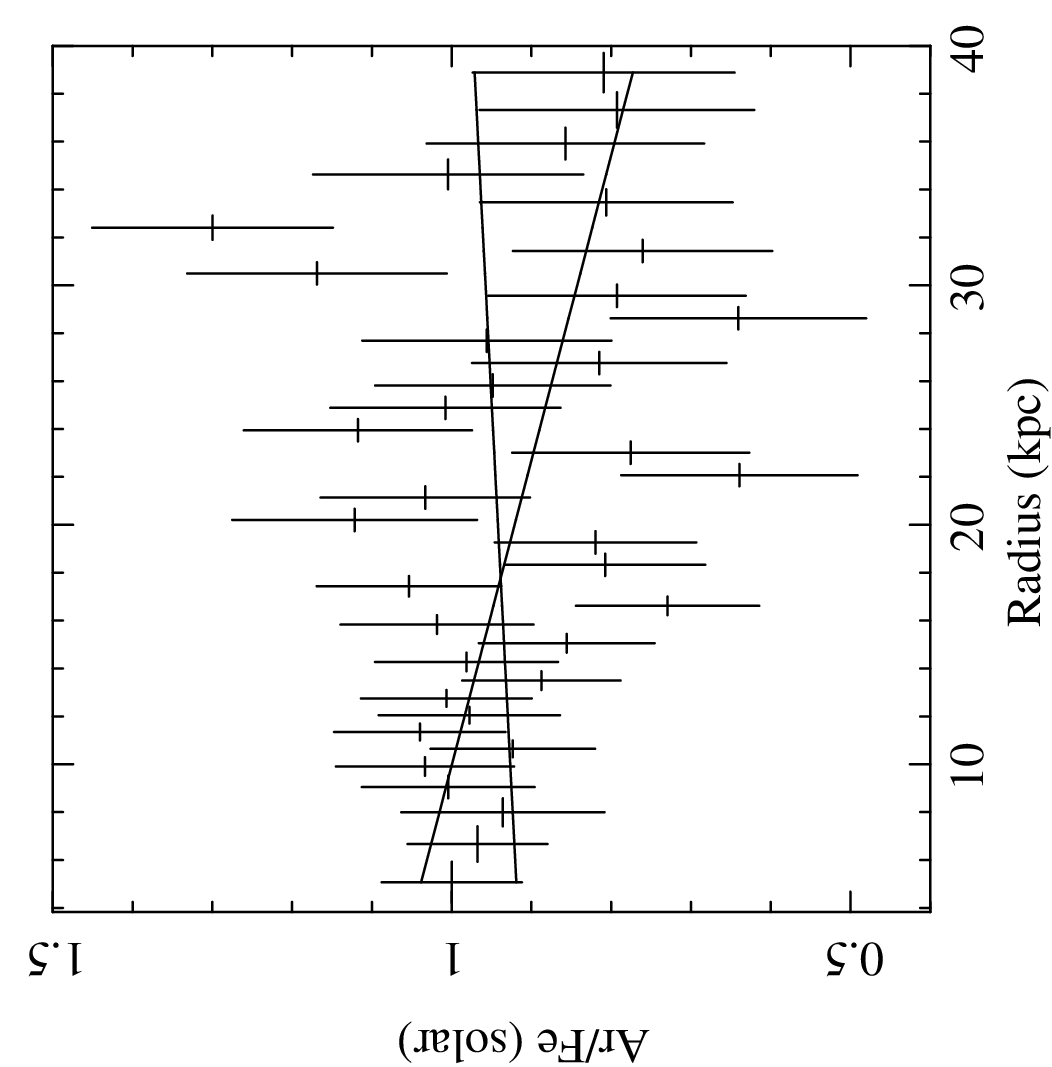}}
\scalebox{0.32}{\includegraphics[angle=270]{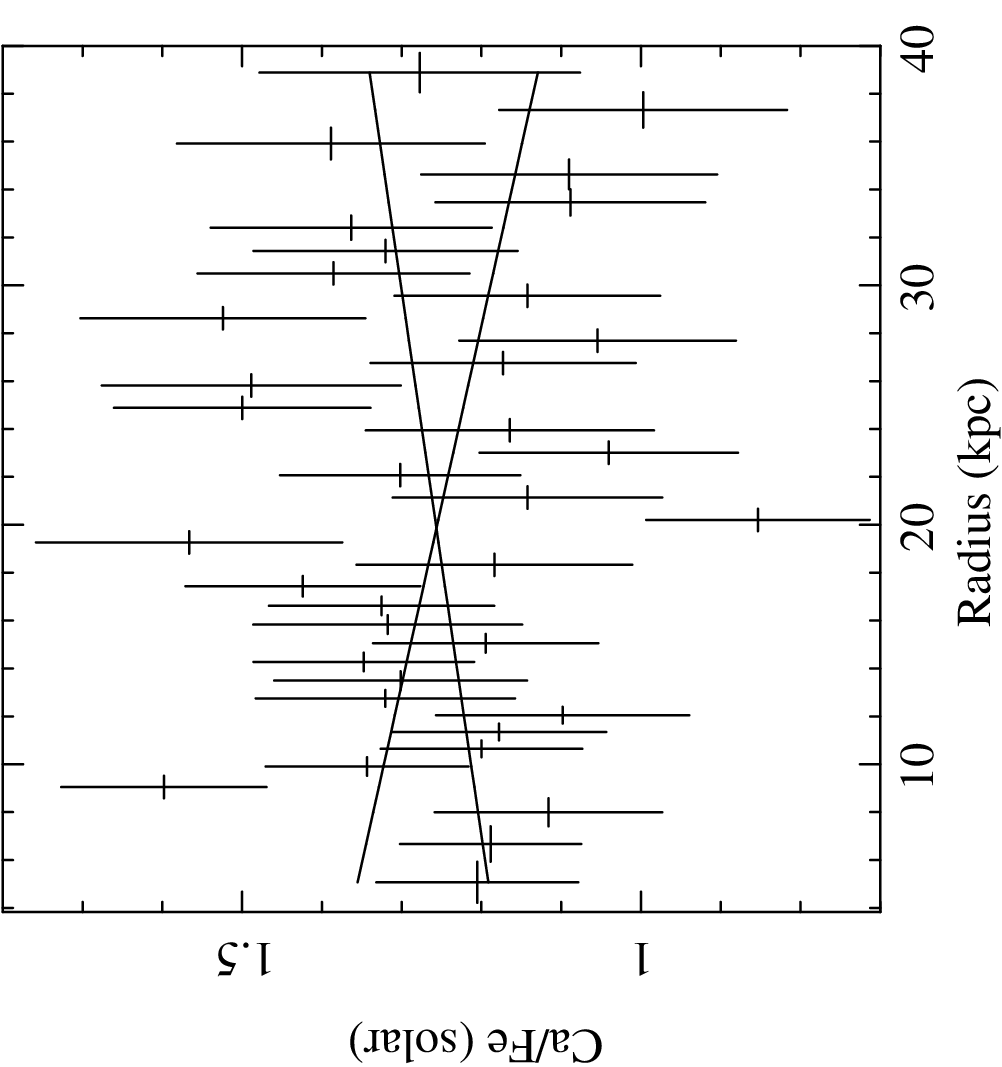}}
\scalebox{0.32}{\includegraphics[angle=270]{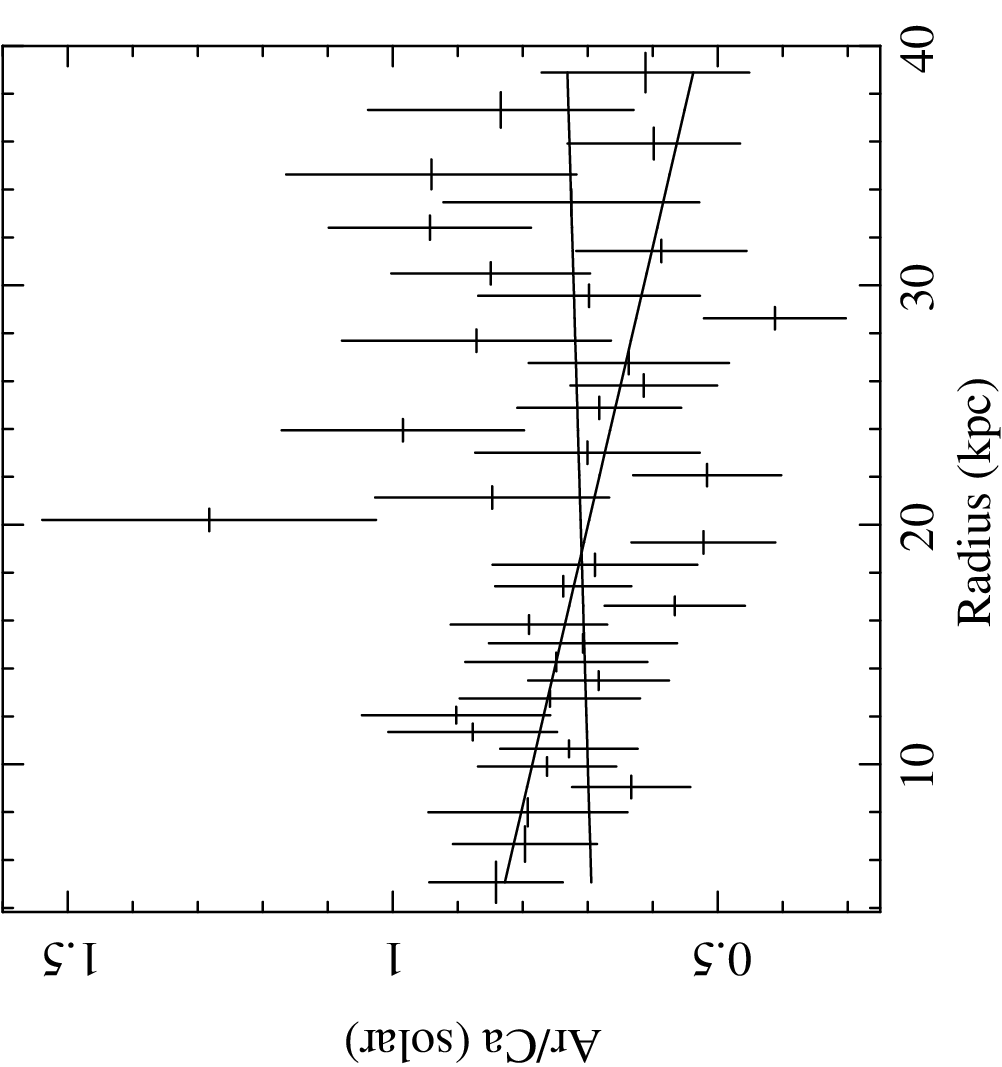}}
\caption{Top row: abundance profiles of Ar (left) and Ca (right). 
Bottom row: abundance ratio profiles of Ar/Fe (left), Ca/Fe (middle), and
Ar/Ca (right). The 95 per cent lower and upper limits of the slopes of the radial distributions of the abundance ratios are overplotted. 
}
\label{fig:ArCa}
\end{figure*}

\begin{figure*}
\scalebox{0.32}{\includegraphics[angle=270]{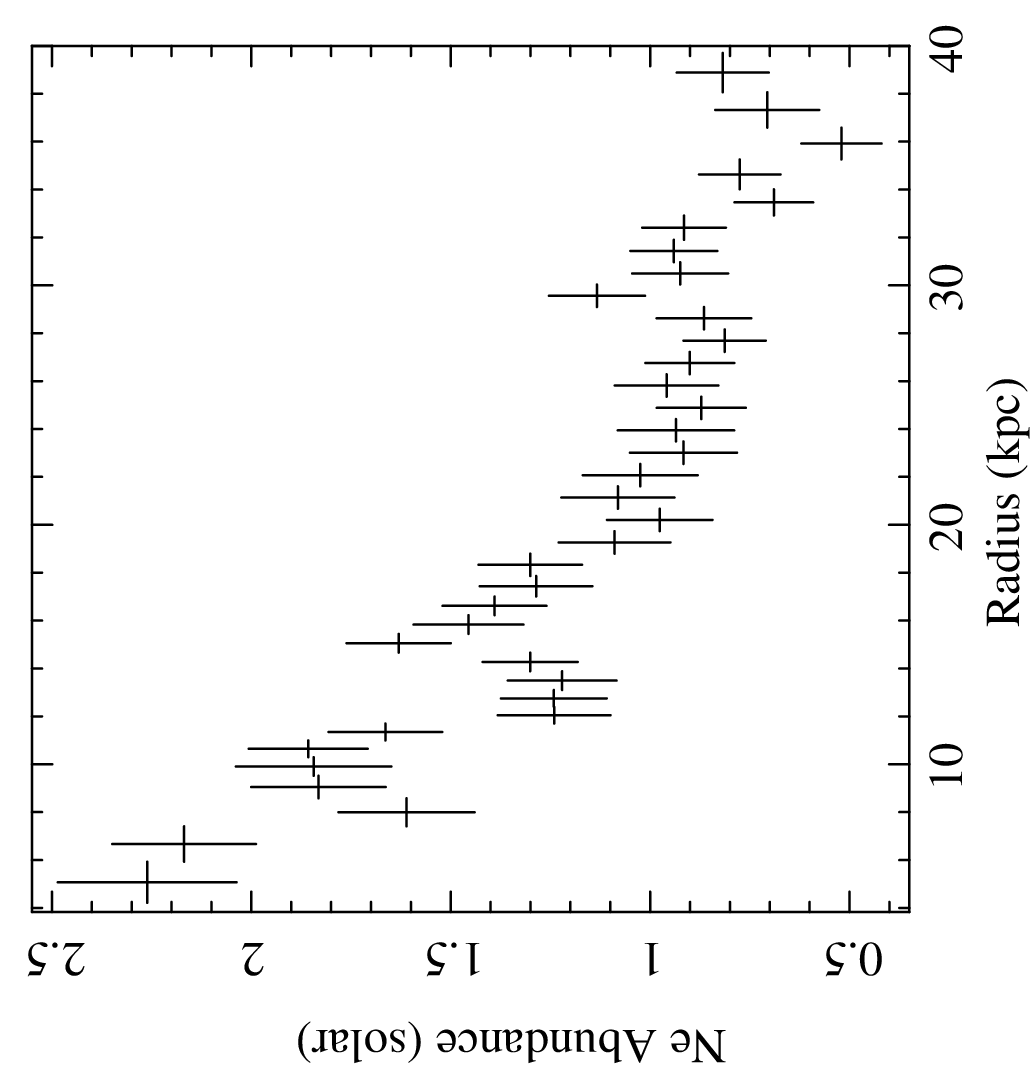}}
\scalebox{0.32}{\includegraphics[angle=270]{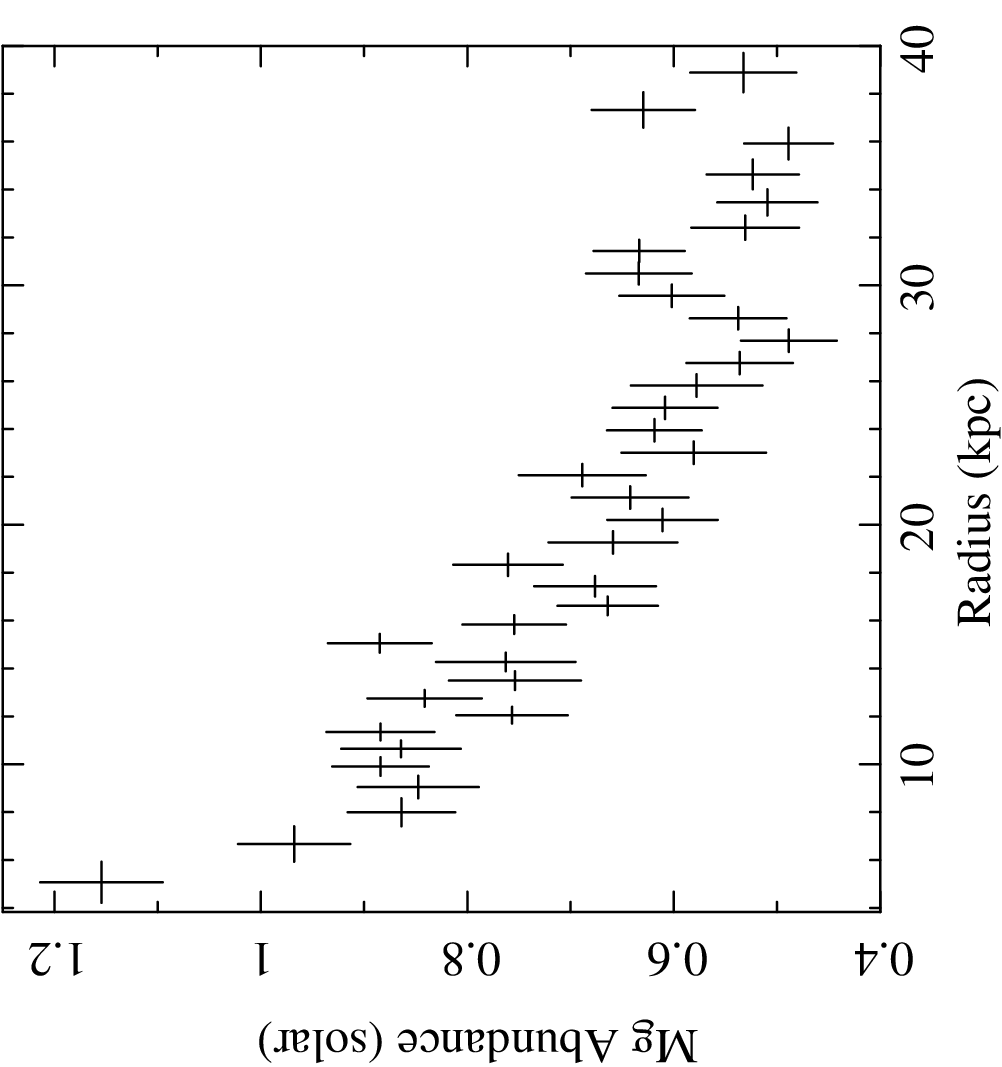}}
\scalebox{0.32}{\includegraphics[angle=270]{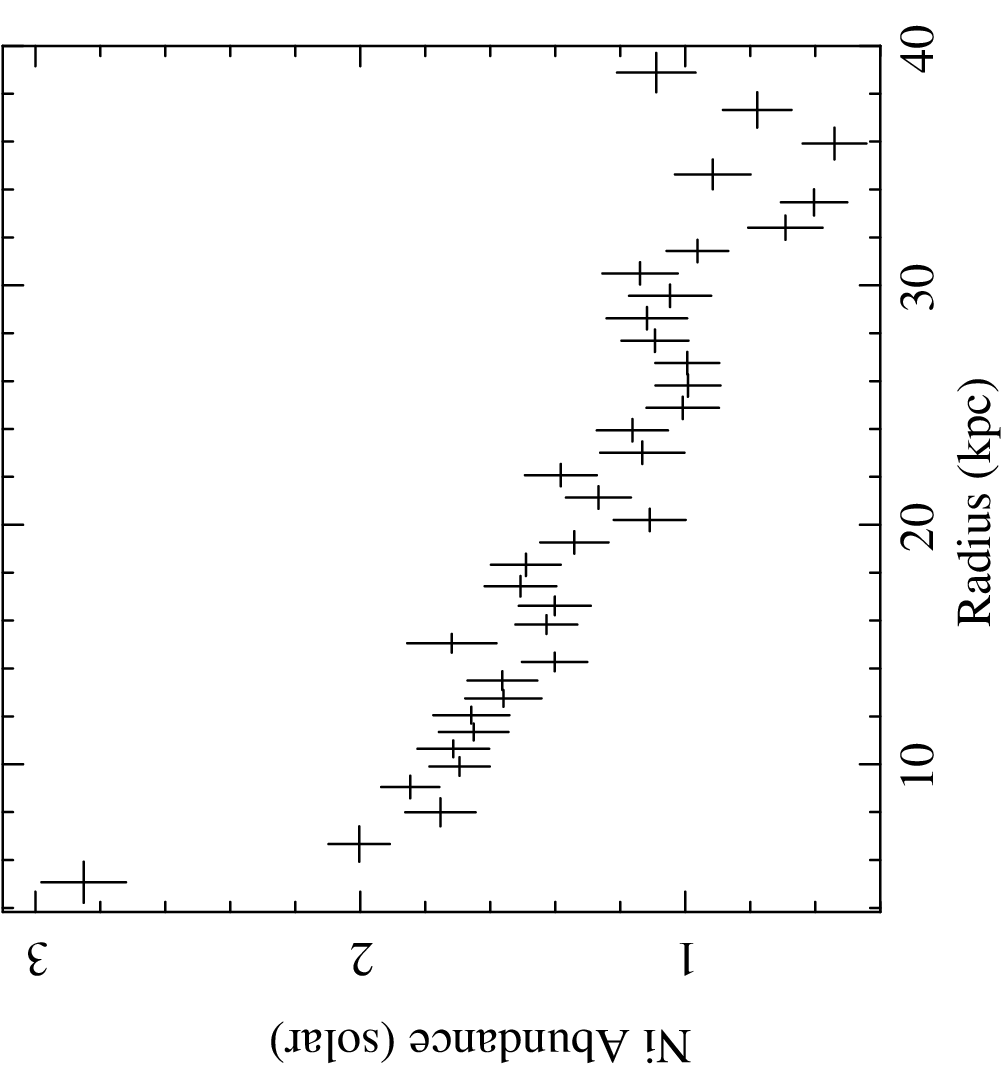}}
\scalebox{0.32}{\includegraphics[angle=270]{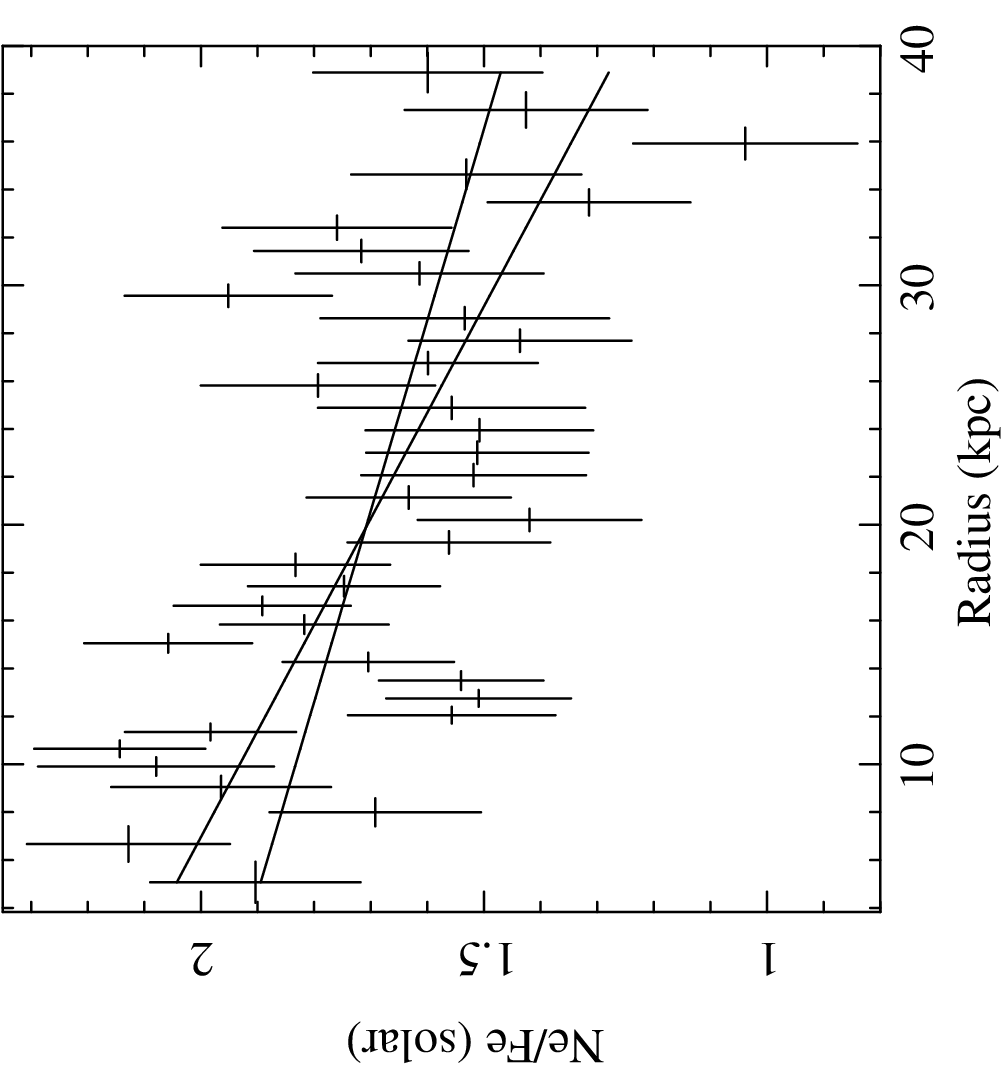}}
\scalebox{0.32}{\includegraphics[angle=270]{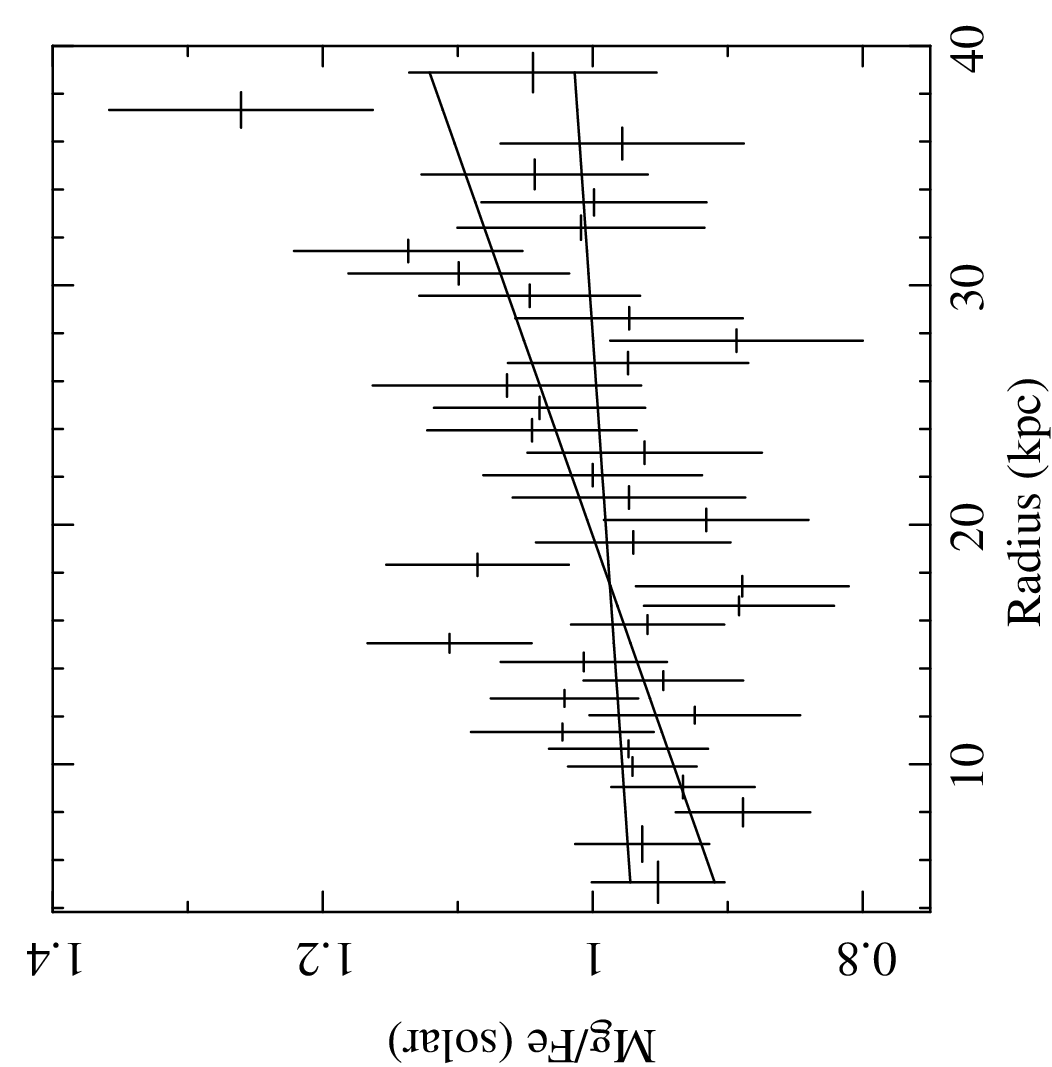}}
\scalebox{0.32}{\includegraphics[angle=270]{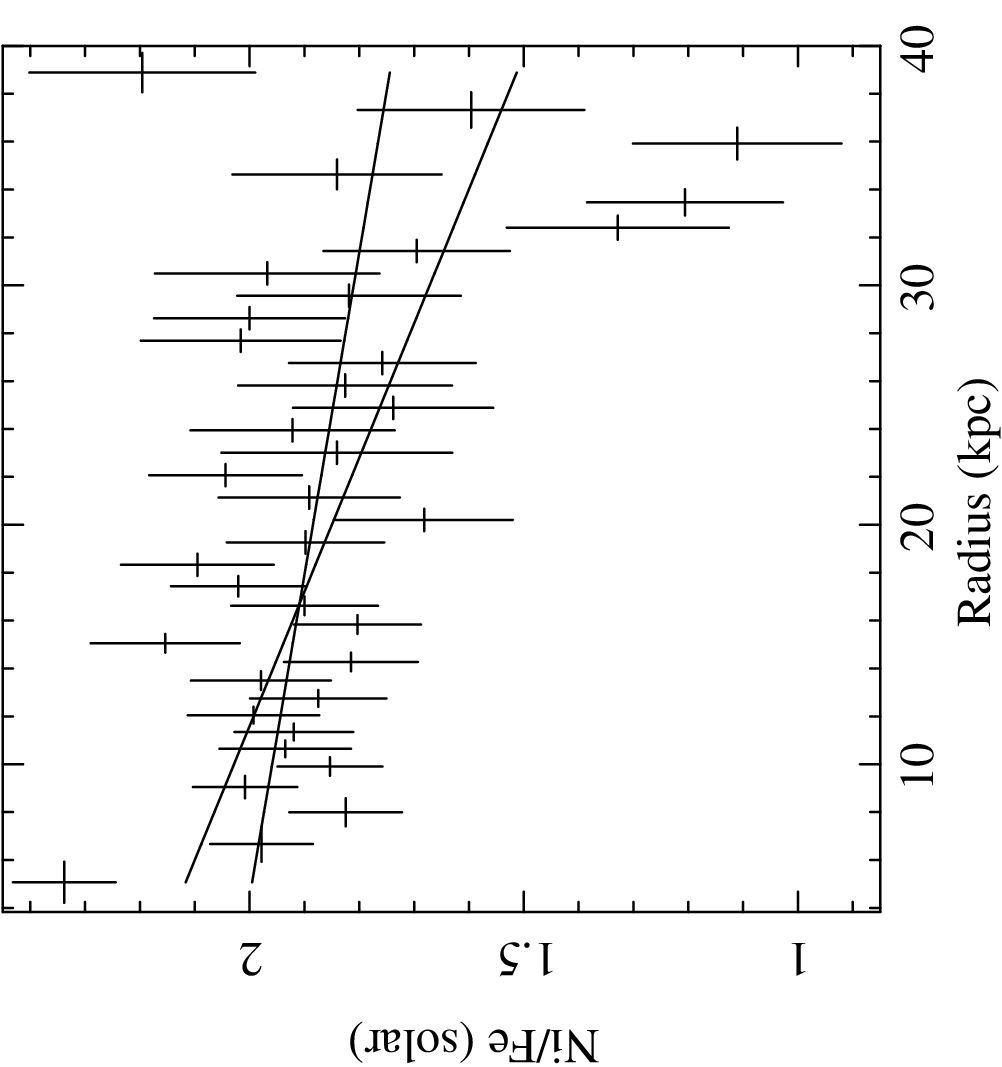}}
\caption{Top row: abundance profiles of Ne (left), Mg (middle), and
Ni (right). Bottom row: abundance ratio profiles of Ne/Fe (left), 
Mg/Fe (middle), and Ni/Fe (right). The Ne, Mg, and Ni lines are blended with the Fe-L line emission. Therefore,
significant modeling bias may be present in the determination of
the abundances of these elements. 
We note, however,
that the abundances of Ne, Mg, and Ni are all centrally peaked,
regardless of the choice of plasma code. 
The 95 per cent lower and upper limits of the slopes of the radial distributions of the abundance ratios are overplotted. 
}
\label{fig:NeMgNi}
\end{figure*}

\begin{table}
\begin{center}
\caption{Summary of the results of the measured radial profiles of the abundance ratios in the ICM.
The data were fit in the 4--40~kpc radial range to a linear model of the form $Z=a+b r$, where $Z$ is
the abundance ratio and $r$ is the radius in kpc.
Columns note the specific abundance ratio, best fit parameters
$a$ (in Solar units) and $b$ 
(in units of $10^{-3}$ Solar kpc$^{-1}$), and the $\chi^2/\nu$.
In cases where the {\small APEC} and {\small MEKAL}
plasma codes disagree, we have included the best fit parameters for each
code. 
The errors on parameter $a$ are at the 68 per cent confidence level. The errors on the slope $b$ are at the 68 per cent and at the 95 per cent (in the parenthesis) confidence level. 
The 95 per cent confidence upper and lower limits of the slope $b$ are overplotted on each abundance ratio profile
(Figs. \ref{fig:FeSiS}d-f, \ref{fig:ArCa}c-e, \ref{fig:NeMgNi}d-f).
We note that only the Ni/Fe abundance ratio profile is not fit well by a simple
linear model.
}
\label{table:trends}
\begin{tabular}{ c c c c }
Ratio & $a$ & $b$ & $\chi^2/\nu$\\
\hline
Si/Fe & $1.316\pm0.013$ & $-7.5\pm0.7$ (1.3) & $18.4/34$ \\
S/Fe & $1.271\pm0.019$ & $-5.6\pm1.0$ (2.0) & $39.7/34$ \\
Si/S & $1.026\pm0.019$ & $-1.5\pm1.0$ (2.0) & $43.4/34$ \\
Ar/Fe & $1.00\pm0.05$ & $-3.1\pm2.3$ (4.7) & $34.3/34$ \\
Ca/Fe & $1.27\pm0.06$ & $-1.1\pm2.8$ (5.5) & $40.1/34$ \\
Ar/Ca & $0.78\pm0.05$ & $-3.7\pm2.4$ (4.8) & $36.2/34$ \\
Ne/Fe$_{\rm APEC}$ & $2.02\pm0.07$ & $-15.9\pm3.3$ (6.7) & $42.3/34$ \\
Ne/Fe$_{\rm MEKAL}$ & $0.80\pm0.04$ & $-4.4\pm1.9$ (3.8) & $43.9/34$ \\
Mg/Fe & $0.92\pm0.02$ & $+3.7\pm1.3$ (2.5) & $26.5/34$ \\
Ni/Fe$_{\rm APEC}$ & $2.12\pm0.05$ & $-12.6\pm2.6 (5.2)$ & $65.9/34$ \\
Ni/Fe$_{\rm MEKAL}$ & $2.13\pm0.05$ & $-19.0\pm2.5$ (5.1) & $56.9/34$ \\
\end{tabular}
\end{center}
\end{table}

We determine the abundance profiles for Fe, Si, S, Ar, Ca, Ne, Mg, and Ni and find that all elements have a centrally peaked distribution.
We also measure the abundance ratios of the individual elements with respect to Fe. In the 4--40~kpc radial range, we fit a linear model of the form
$Z=a+b r$ to these measurements, where $Z$ is the abundance ratio, 
$r$ is the radius in kpc, $a$ is the normalization
of the linear trend, and $b$ is the slope.
Table \ref{table:trends} summarizes the best-fit linear relations for each
of the abundance ratio profiles. 
For the Ne/Fe and Ni/Fe ratios, which have significant systematic uncertainties, we
present the best-fit parameters for profiles determined using both the {\small APEC} and {\small MEKAL}  plasma codes.

The absolute abundances determined by the {\small MEKAL} plasma code are larger by 1--2 per cent for Si, by $\sim$3 per cent for S, by $\sim$15 per cent for Ar and Ca, and by $\sim$20 per cent for Fe than the values obtained using {\small APEC}. The slopes of the abundance ratio profiles for these
elements are, however, consistent with those determined by the {\small APEC} code. 
The abundances of Ne, Ni, and Mg are significantly different when determined by the {\small
MEKAL} plasma code. A discussion of modeling biases for the abundances of these elements
can be found in Sect.~\ref{section:bias}. 

For gas with $kT\approxgt2.0$ keV, the O abundance is extremely difficult to 
determine with {\it Chandra}.
At the \ion{O}{viii} line energy of 0.65~keV the effective area of the ACIS detectors is significantly affected 
by buildup of contamination. Additionally, in the lower surface brightness areas at larger radii the O abundance measurements are sensitive to the assumed Galactic foreground model. 
Because our O abundance measurements have large systematic uncertainties and may be strongly biased, we do not report their best fit values. 

We have examined the abundance ratios separately to
the north and south of M\,87. Although the overall abundances are larger to the
south (see Paper I), the abundance ratios determined from the northern
and southern sectors agree well with each other and the azimuthally-averaged
analysis presented here.

\subsection{Fe, Si, and S abundances}
\label{section:SiS}

The top row of Fig. \ref{fig:FeSiS} shows the Fe, Si, and S abundance profiles, respectively. 
The bottom row shows the abundance ratio profiles of
Si/Fe, S/Fe, and Si/S. 
We emphasize that these three elements have the 
most robustly determined abundances, with statistical uncertainties of less than 5 per cent.

The Fe abundance profile (Fig. \ref{fig:FeSiS}a)
peaks at $Z_{\rm Fe}>1.2$ Solar in the central regions. 
The Si and S abundance profiles (Fig. \ref{fig:FeSiS}b-c) peak at
a slightly larger central value of $Z_{\rm Si;S}\sim1.5$ Solar. 
These profiles then exhibit steady declines to $\sim0.6$ Solar by $r\sim25$~kpc.

A significant enhancement, or `bump', in the
Fe and Si (and possibly S) abundances is seen at $r\sim$30~kpc.  This is approximately the radius
at which the bright X-ray arms terminate. As discussed
in Paper~I, this bump may be due to the uplift of cool, metal-rich material in the wake of buoyantly rising radio bubbles.

Most importantly, we observe, for the first time, a radially decreasing trend in the
Si/Fe and S/Fe abundance ratios (Fig. \ref{fig:FeSiS}d-e).
They both peak at $\sim1.3$ Solar near the core and decline to $\sim1.0$ Solar
by $r\sim35$~kpc. 
Both the Si/Fe and S/Fe profiles are well described by 
a linear decline as a function of radius.
As seen in Table \ref{table:trends}, both profiles have similar slopes 
to one another and are significantly steeper than a constant ratio (at the 5--11$\sigma$ level).
The Si/Fe and S/Fe ratios of $\sim$0.9~Solar measured at radii between 50--200~kpc with {\it Suzaku} (Simionescu et al. 2010) indicates that 
beyond $r\sim35$~kpc these abundance ratio profiles flatten. 

The Si/S abundance ratio profile (Fig. \ref{fig:FeSiS}f) is an interesting cross check upon 
the determination of these abundances. Because both elements are created
in roughly equal quantities by both ${\rm SN_{CC}}$ and ${\rm SN\;Ia}$
(see Iwamoto et al. 1999; Nomoto et al. 2006),
the Si/S abundance ratio is insensitive to the relative fraction of 
supernovae that explode as ${\rm SN_{CC}}$ and ${\rm SN\;Ia}$. Any trend
in the Si/S abundance ratio profile is instead primarily affected by a 
change in the average yields of ${\rm SN_{CC}}$ and ${\rm SN\;Ia}$ as
a function of radius. The Si/S abundance ratio is consistent with being flat as a function of radius and 
its observed mean value is $Z_{\rm Si}/Z_{\rm S}=1.002\pm0.009$ Solar.

The systematic uncertainties involved in the modelling of Si and S lines are relatively small due to the favorable location of their lines in the 2--3~keV range, where no residual Fe-L emission is present. Possible biases due to multi-temperature structure are minimized by our choice of extraction regions which avoid clear density and temperature inhomogeneities and have an approximately isothermal azimuthal temperature structure (see Paper I). Nonetheless, we performed detailed spectral simulations of multi-temperature plasmas to investigate possible biases due to residual structure and projection effects. We mixed {\small APEC} thermal models with slightly different temperatures and metallicities, where the cooler plasma is more enriched than the hotter one. The simulated gas mixtures span a grid of possible projection effects and unresolved temperature structure. We fitted these simulated spectra with a single temperature model. The potential bias in Si/Fe, S/Fe and Si/S ratios was always less than 10~per cent. 

Furthermore, we note that systematic uncertainties related to the effective area calibration of the ACIS detectors around the Si edge would be likely to affect the Si and S abundance profiles differently. Therefore, the striking similarity of the Si/Fe and S/Fe profiles strongly indicates that the detection of these trends is robust.

\subsection{Ar and Ca abundances}
\label{section:Argon}

The top row of Fig. \ref{fig:ArCa} shows the Ar (left) and Ca (right) abundance profiles. The bottom row shows the radial distributions of the
 Ar/Fe (left), Ca/Fe (middle), and Ar/Ca (right) abundance ratios.
Because the Ar and Ca lines have small equivalent widths, their abundances
are more difficult to measure precisely than Fe, Si, and S:
our data enable us to measure the Ar and Ca abundances to within 20 per cent statistical uncertainty.
Our spectral
fits reveal that the Ar and Ca abundances are relatively insensitive to small changes in the
temperature and Fe abundance. Therefore, systematic uncertainties on these measurements are likely small.

The Ar and Ca abundance profiles (Fig. \ref{fig:ArCa}a-b) are also centrally peaked.
The Ar and Ca abundances peak at $Z_{\rm Ar}\sim1.2$ and $Z_{\rm Ca}\sim1.4$ Solar, respectively.
They decrease to $Z_{\rm Ar}\sim0.5$ and $Z_{\rm Ca}\sim0.6$ Solar respectively by $r\sim35$~kpc. 
As with Fe and Si, both profiles show a marginal, but plausible increase in abundance
at $r\sim30$~kpc that may be due to the uplift of cool, metal-rich material, as described in Paper I. 

Both the Ar/Fe and Ca/Fe abundance ratio profiles 
(Fig. \ref{fig:ArCa}c-d) are consistent with constant values of $Z_{\rm Ar}/Z_{\rm Fe}=0.94\pm0.02$ and $Z_{\rm Ca}/Z_{\rm Fe}=1.25\pm0.03$ Solar, respectively.

Similarly to the Si/S abundance ratio profile, the Ar/Ca abundance ratio profile
(Fig. \ref{fig:ArCa}e)
is also an interesting cross-check upon the determination of 
these abundances. Like Si and S, Ar and Ca are also created in similar quantities 
by ${\rm SN_{CC}}$ and ${\rm SN\;Ia}$. Therefore, the Ar/Ca abundance ratio profile
is insensitive to the relative number of supernovae that explode as ${\rm SN_{CC}}$
and ${\rm SN\;Ia}$. Instead, the Ar/Ca abundance ratio is primarily sensitive
to changes in the average yields of ${\rm SN\;Ia}$. 
The Ar/Ca abundance ratio is consistent with being flat as a function of radius and its observed mean value is 
$Z_{\rm Ar}/Z_{\rm Ca}=0.71\pm0.02$ Solar.

\subsection{Ne, Mg, and Ni abundances}
\label{section:Mg/Fe}

The top row of Fig.~\ref{fig:NeMgNi} shows the abundance profiles of Ne (left), Mg (middle), and Ni (right),
respectively. The bottom row shows the abundance ratio profiles of Ne/Fe (left), Mg/Fe (middle), 
and Ni/Fe (right).
There are significant systematic uncertainties present in the determination of the 
Ne, Mg, and Ni abundances (see Sect.~\ref{section:bias}).

The inferred abundance profiles of Ne, Mg, and Ni are all centrally peaked at 
$Z_{\rm Ne}\sim 2.3$ Solar, $Z_{\rm Mg}\sim1.2$ Solar, and $Z_{\rm Ni}\sim2.8$ Solar, respectively.
The profiles decline to $Z_{\rm Ne}\sim1.0$ Solar, $Z_{\rm Mg}\sim0.6$ Solar, and $Z_{\rm Ni}\sim1.0$ Solar
by $r\sim25$~kpc. 
Although there are differences between the abundances measured with {\small APEC} and {\small MEKAL},
strong central peaks in the abundance profiles of Ne, Mg, and Ni are found with both codes.
These profiles also show the enhancements at $\sim$30~kpc. 
The Ne/Fe abundance ratio determined by both plasma codes is centrally peaked.

The Mg/Fe ratio is marginally consistent with being flat as a function of radius and its mean values is
$Z_{\rm Mg}/Z_{\rm Fe}=0.988\pm0.012$ Solar. This is in agreement with previous results once the differences
between plasma codes are taken into account.
(see \eg Matsushita \etal 2003, Simionescu \etal 2010). 

The Ni/Fe results derived using both plasma codes are
poorly fit to a linear model. The reduced $\chi^2$s are high, $\chi^2/\nu\sim60/34$.
Primarily this is due to increased scatter in the Ni measurements at large radii ($r>30$~kpc).
Within the radial range of $5<r<30$~kpc and using the {\small APEC} code
the Ni/Fe abundance ratio is fit well by a single, constant value of $Z_{\rm Ni}/Z_{\rm Fe}=1.92\pm0.02$ Solar.
However, a large drop in the Ni/Fe abundance ratio at radii $r\ge30$~kpc is reproduced by both plasma codes. 

\subsubsection{Modelling uncertainties in Ne, Mg, and Ni measurements}
\label{section:bias}

The spectral resolution of the CCD type detectors on {\it Chandra} does not allow us to resolve
the individual Ne and Ni lines within the Fe-L complex. 
The Mg lines are also blended with Fe-L line emission.
Therefore, the Ne, Mg, and Ni abundances dependent critically on
the modeling of the Fe-L complex.

The Ne, Mg, and Ni abundances determined by the {\small MEKAL}
and {\small APEC} plasma codes disagree, with the {\small MEKAL} Mg abundances being a factor of two lower 
than the values determined with {\small APEC}.
Both plasma models remain incomplete and therefore the results must be treated with caution.

Previous results from the Reflective Grating Spectrometer (RGS) aboard {\it XMM-Newton}, which can resolve separately the Ne lines within the Fe-L complex,
reveal that the Ne/Fe abundance ratio is above the Solar value (see Werner \etal 2006b), consistent with our Ne abundance measurement. 

Incorrect ionization balance for Ni in the plasma models can lead to temperature dependent biases in its abundance determination (see B\"ohringer \& Werner 2010). At $kT\sim2.1$ keV, the Ni abundances determined with {\small APEC} are 20 per cent lower than those determined using {\small MEKAL}. 
This difference decreases linearly with temperature.
At $kT\sim2.7$ keV, the best fit Ni abundances determined using the two plasma codes are consistent. Our Ni abundances are determined from their L-shell lines. 
Using additional spectral data between $7.0-8.0$ keV to include the K-shell lines does not improve our determination of the Ni abundance.

\section{Implications for supernova enrichment models}
\label{section:decline}

The elements observed in the X-ray emitting gas of clusters of galaxies reveal the integrated chemical enrichment history by supernovae.
The standard picture of chemical enrichment in clusters of galaxies for at least the past 12 years has been an
early enrichment by ${\rm SN_{CC}}$ products that are now well mixed in the ICM, in combination with a later
contribution by ${\rm SN\;Ia}$, which have longer delay times and primarily enrich the region surrounding the cD galaxy.
Our data clearly disagree with this picture.  Both Si and S show central
abundance peaks that are larger than that of Fe (see Figs. \ref{fig:FeSiS}e-f,
\ref{fig:NeMgNi}d). The abundances of other elements (Ar, Ca, Ne, and Mg) 
show central abundance peaks as well (see Figs. \ref{fig:ArCa}c-d, \ref{fig:NeMgNi}e). 
These results force us to rethink our models of the
chemical enrichment in clusters of galaxies. 

Here we discuss possible chemical enrichment scenarios that may 
explain the observed abundance ratio profiles.
These include centrally peaked ${\rm SN_{CC}}$ products due to stellar winds, intermittent star formation, and early enrichment of the low entropy gas. We also discuss how  radial changes in the stellar initial mass function, and pre-enrichment of ${\rm SN_{CC}}$ progenitors, or a diversity 
in the population of ${\rm SN\;Ia}$ can affect the observed distributions of metals.

\begin{figure*}
\scalebox{0.32}{\includegraphics[angle=270]{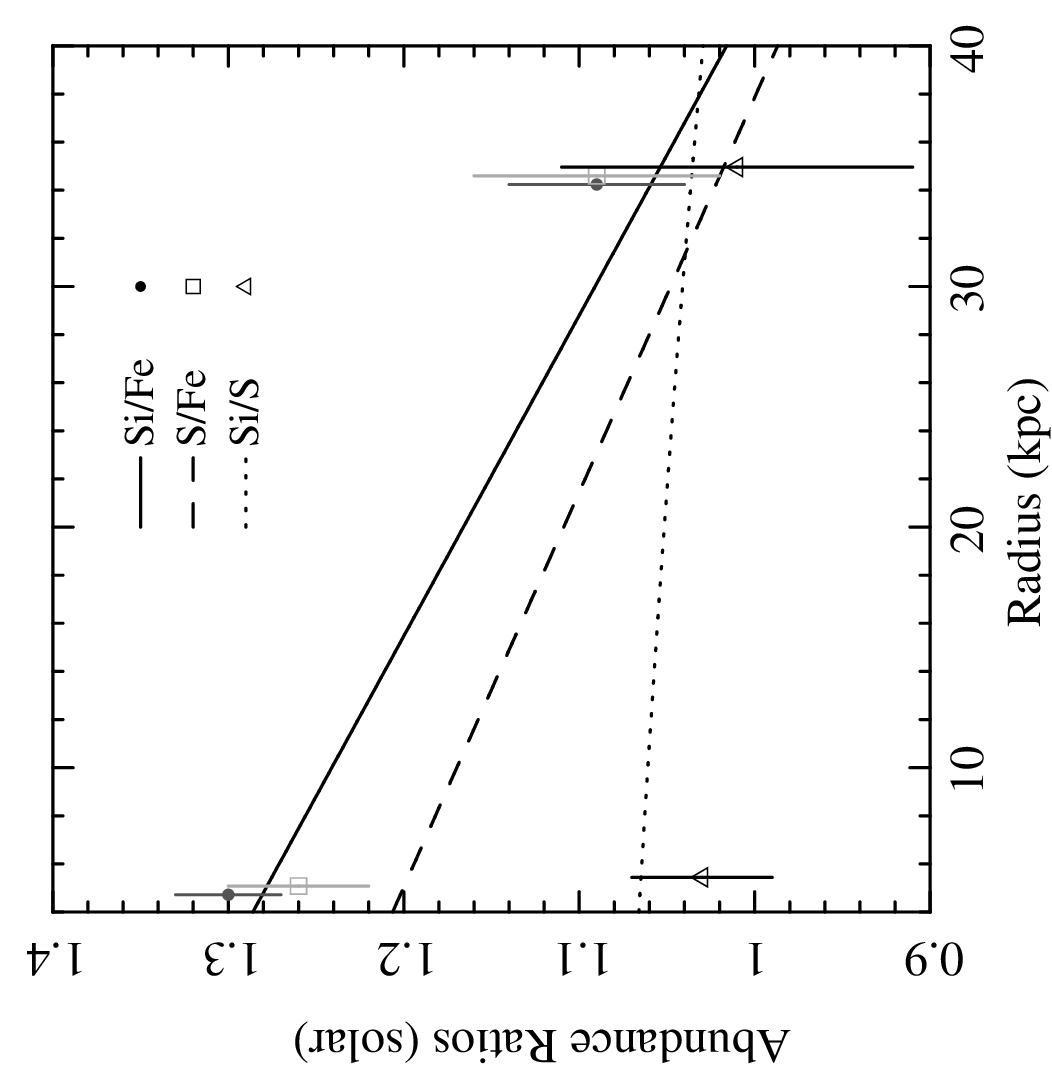}}
\scalebox{0.32}{\includegraphics[angle=270]{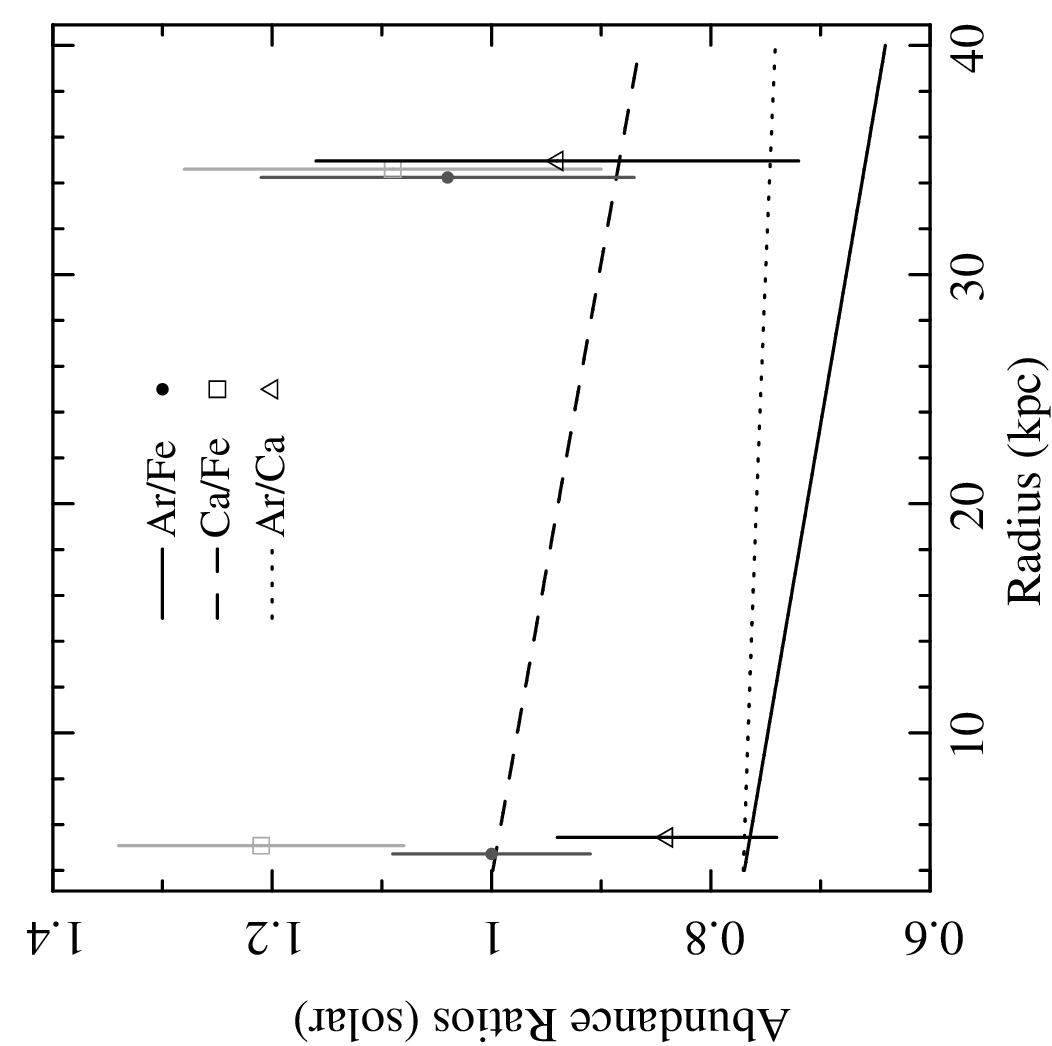}}
\scalebox{0.32}{\includegraphics[angle=270]{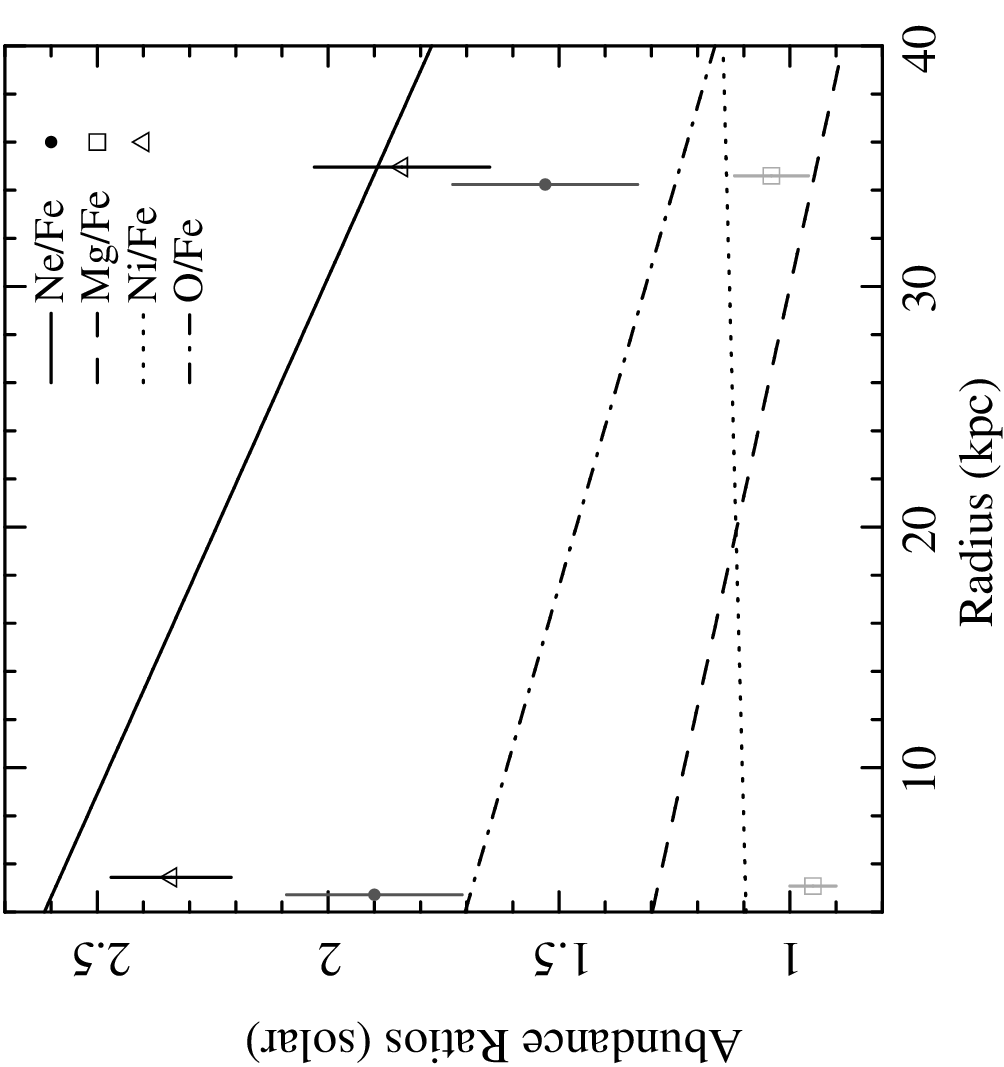}}
\caption{Predicted abundance ratio profiles using the calculated fraction of ${\rm SN\;Ia}$, $f(r)$, from the Si/Fe abundance ratio profile with the measured data points at 4~kpc and 35 kpc over-plotted. The predicted abundance
ratios of S/Fe and Si/S agree well with observations. However, the predicted radial distributions of the abundance ratios of the other elements do not match the observations well. }
\label{fig:sncc_peak}
\end{figure*}

\subsection{Centrally peaked ${\rm SN_{CC}}$ products}
\label{section:sncc}

The most straight forward interpretation of the centrally peaked Si/Fe and S/Fe abundance ratio profiles, 
is that there is a dominant contribution of SN$_{\rm CC}$ products to the enrichment of the lowest entropy gas in the central regions of the galaxy. 
The contribution of SN~Ia products rises with increasing radius, out to $r\sim 35$~kpc. 

We calculate the fraction of all supernovae that explode as ${\rm SN\;Ia}$ as a function of radius using the equation
\begin{equation}
\left(\frac{M_{\rm Z}}{M_{\rm Fe}}\right)_{\rm obs}=\frac{f M_{\rm Z; SNIa}
+(1-f) M_{\rm Z; SN_{\rm CC}}}{f M_{\rm Fe; SNIa} + (1-f) M_{\rm Fe; SN_{\rm CC}}},
\end{equation}

\noindent where $\left(M_{\rm Z}/M_{\rm Fe}\right)_{\rm obs}$ 
is the mass ratio of element Z with respect to Fe converted from our abundance measurements 
using the proto-Solar values by Lodders (2003), $f=N_{\rm SNIa}/(N_{\rm SNIa}+N_{\rm SN_{CC}})$ 
is the fraction of all supernovae that explode as ${\rm SN\;Ia}$, and $M_{\rm Z; SNIa}$ and $M_{\rm Z; SN_{\rm CC}}$ are the average
theoretical yields for ${\rm SN\;Ia}$ and ${\rm SN_{CC}}$, respectively. For SN~Ia, we use the yields of the WDD2 delayed-detonation model
by Iwamoto \etal (1999), except where explicitly stated.
Because Si is the second most reliably measured chemical element after Fe, we calculate the number fraction of ${\rm SN\;Ia}$, $f(r)$, 
using the Si/Fe ratio. 

The average theoretical yields of ${\rm SN_{CC}}$ are calculated using the values by Nomoto \etal (2006) under the
assumption of the Salpeter initial mass function (IMF). In detail,
\begin{equation}
M_{\rm Z;SN_{CC}}=\frac{\int_{10 M_\odot}^{50 M_\odot}\; M_{\rm Z}(M) M^{-\alpha} dM}
{\int_{10 M_\odot}^{50 M_\odot}\; M^{-\alpha} dM},
\end{equation}
\noindent where $M_{\rm Z;SN_{CC}}$ is the theoretical yield 
of element Z for the weighted-average of core-collapse supernovae, $M_{\rm Z}(M)$ are the atomic yields as a 
function of stellar mass for core-collapse supernovae, and $\alpha=2.35$
for the Salpeter IMF.

We calculate a central fraction of ${\rm SN\;Ia}$ of $f=0.10$ at radius $r\sim50$
arcsec (3.9~kpc). This fraction rises to $f=0.19$ at radius $r\sim500$ arcsec (38.8~kpc).
Based on these fractions, we can predict the radial profiles of the abundance ratios for other elements. The results are shown in 
Fig. \ref{fig:sncc_peak}.

The S/Fe and Si/S abundance ratio profiles predicted by the central 
increase in the relative fraction of ${\rm SN_{CC}}$ agree well with
our observed profiles.
While the observed slopes of the Ca/Fe and Ar/Fe profiles are also consistent with the centrally peaked SN$_{\rm CC}$ enrichment, their normalizations are higher than those predicted by the model. Our Ca/Fe abundance ratio is, however, consistent with the average value measured in 22 clusters of galaxies by de Plaa et al. (2007) once systematic differences in the plasma codes used are taken into account.  These authors showed that the measured abundance ratios can be well fit with a one-dimensional delayed-detonation SN Ia model, calculated on a grid introduced in Badenes et al. (2003), with a deflagration to detonation density of $2.2\times10^7$~g~cm$^{-3}$ and kinetic energy of $1.16\times10^{51}$~ergs, which was shown to fit best the properties of the Tycho supernova remnant (Badenes et al. 2006).
 
The predicted Mg/Fe abundance ratio profile disagrees with the measurements. The Mg abundance measurements are, however, affected by systematic uncertainties arising from the incomplete modeling of the Fe-L shell transitions. 

\subsubsection{Centrally peaked SN$_{\rm CC}$ products due to intermittent star-formation}

If both the low entropy gas in the cluster core and the bulk of the surrounding ICM were enriched around the same time, with a similar mixture of SN$_{\rm CC}$ and SN~Ia, then the expected Si/Fe and S/Fe ratios would be constant with radius.  
However, out to a radius of $r=35$~kpc, the observed Si/Fe and S/Fe ratios decreases from $\sim$1.3 to $\sim$1.0 Solar, before staying flat out to at least a radius of 200~kpc (Simionescu et al. 2010). In order to produce the observed Si/Fe peak on top of this flat profile, an additional $1.6\times10^7$~SN$_{\rm CC}$ would have to explode in the centre of M~87. This is about 10 per~cent of all SN$_{\rm CC}$ that are required for the chemical enrichment of the innermost $r<40$~kpc region. Assuming a Salpeter IMF, such number of supernovae require $\sim10^9$~M$_{\odot}$ of star formation. Because of the continuing production of Fe by SN~Ia in the cD galaxy, this value is a lower limit. The SN$_{\rm CC}$ enrichment would also have to occur on a time scale shorter than the production of Fe by SN~Ia. While, there is no evidence for current star formation in M~87, and the 95 per cent confidence upper limit on radiative cooling from the coolest ICM phase is 0.06~M$_{\odot}$~yr$^{-1}$ (Werner et al. 2010),  we can not rule out past intermittent star forming episodes in the central regions of the galaxy. Star formation, at rates reaching several tens and up to hundreds of solar masses per year, has been observed in the brightest cluster galaxies of some other cooling core clusters (e.g. McNamara et al. 2006, Ogrean et al. 2009, Ehlert et al. 2011). Therefore, several intermittent star forming episodes at $\sim$10~M$_{\odot}$~yr$^{-1}$ lasting for a total of $\sim2\times10^8$~yr would not be surprising. 

\subsubsection{Centrally peaked SN$_{\rm CC}$ products due to stellar winds}

Stellar mass loss is an important source of metals in the hot gas surrounding giant elliptical galaxies like M\, 87. The material that formed the current stellar population of M\, 87 was most likely pre-enriched primarily by SN$_{\rm CC}$ products. Some of these metals are then returned into the ICM/ISM by stellar winds.  Assuming a single-age passively evolving stellar population with a Salpeter initial mass function, Ciotti et al. (1991) predict a stellar mass loss rate of  
\begin{equation}
\dot{M}_{\star}({\rm t})\approx 1.5\times10^{-11} L_{\rm B} t_{15}^{-1.3},
\end{equation}
where $L_{\rm B}$ is the present-day B-band luminosity in units of $L_{{\rm B}\odot}$ ($L_{\rm B} = 8.4\times10^{10} L_{{\rm B}\odot}$ for M\, 87; Gavazzi et al. 2005) and $t_{15}$ is the age of the stellar population in units of 15~Gyr (the formula is valid in the range from $\sim$0.5 to over 15 Gyr). Integrating this equation under the assumption that the current stellar population is 10.5 Gyr old (formed at $z=2.1$), we obtain a gas mass contribution from stellar winds of $1\times10^{11}~M_{\odot}$ during the past 10 Gyr. Assuming that the material from which the current stellar population of M~87 formed was pre-enriched to conservatively $Z_{\rm Si}\sim0.5$ Solar, with a Si/Fe abundance ratio of 1.5 Solar, the total mass of Si returned to the ISM/ICM by the stellar winds is $4.3\times10^7~M_{\odot}$. The total mass of Fe returned is $4.8\times10^7~M_{\odot}$. Under these assumptions, the total mass of Si produced by stellar winds in excess of a Si/Fe ratio of 1 Solar is $1.4\times10^7~M_{\odot}$. The observed total Si mass in excess to a flat profile with Si/Fe=1 Solar around M\, 87 is $3.6\times10^6~M_{\odot}$.  
Therefore, despite all the uncertainties in the estimates of the metal mass loss, the excess Si observed around M~87 could most likely be produced by stellar winds.
Because the initial starbursts, that enriched the material from which the current stellar population formed,  produced predominantly SN$_{\rm CC}$ products, and the Mg$_2$ index indicates that the stellar population of M\, 87 is enriched to more that 1 Solar (Kobayashi \& Arimoto 1999, Matsushita et al. 2003) this scenario can plausibly produce centrally peaked abundance profiles of SN$_{\rm CC}$ products.

\subsubsection{Centrally peaked SN$_{\rm CC}$ products due to strong early enrichment of the low entropy gas}

Another possible mechanism contributing to the observed centrally peaked distribution of SN$_{\rm CC}$ products is strong early enrichment of the lowest entropy X-ray emitting gas and inefficient mixing of this material with the surrounding ICM. If the lowest entropy X-ray emitting gas currently seen in the Virgo Cluster core was at high redshift located in the environments of massive galaxies during their epoch of maximum star formation, then this material may have become enriched significantly by SN$_{\rm CC}$ products. As the cluster formed, this low entropy SN$_{\rm CC}$ enriched gas will naturally sink and accumulate at the base of the cluster potential. Assuming that it does not become well mixed with the surrounding ICM as it does so and it does not cool out of the hot phase, this may lead to the observed peak in metal abundances.

\subsection{Other mechanisms affecting the radial profiles of abundance ratios}
We explore other possible explanations for the observed abundance ratio profiles. Each of these mechanisms predicts a central enhancement in the Si/Fe, S/Fe, Ar/Fe, and Ca/Fe ratios and each of them might, to some degree, contribute to the observed radial trends.

\subsubsection{Changing IMF as a function of radius}
\label{section:IMF}

\begin{figure}
\scalebox{0.43}{\includegraphics[angle=270]{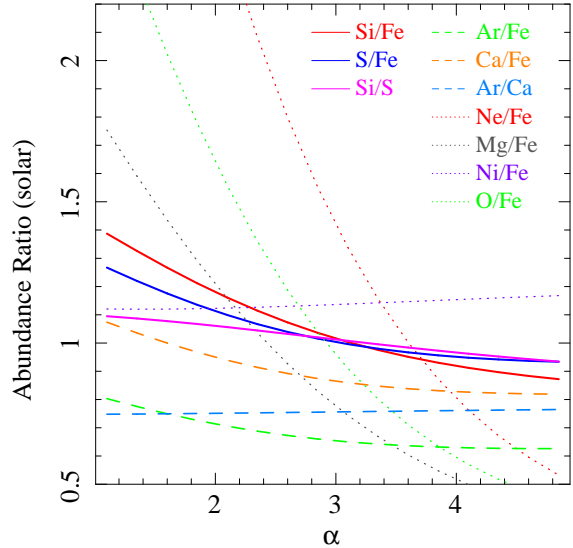}}
\caption{Predicted abundance ratios as a function of the slope
of the stellar initial mass function. The value 2.35 corresponds to the 
Salpeter value. Flatter mass functions, i.e. larger relative fractions of
massive stars exploding as ${\rm SN_{CC}}$, predict a rise in Si/Fe and S/Fe ratios as we observe in the centre of the galaxy. 
Based on ${\rm SN_{CC}}$ yields by Nomoto \etal (2006) and assuming that SN~Ia (WDD2 model by Iwamoto et al. 1999) make up 15 per cent of all supernovae.
}
\label{fig:alpha}
\end{figure}

A stellar IMF, which is flatter (has a smaller $\alpha$ in Equation 2, therefore producing more massive stars) in the central regions of the cluster would produce a central increase of Si-group elements.  
The ratios of chemical elements produced by ${\rm SN_{CC}}$ are a strong function of the mass of the progenitor. 
To examine the effect of the IMF on the predicted abundance ratios, we vary its slope $\alpha$.

Fig. \ref{fig:alpha} shows the predicted abundance ratios assuming a steepening IMF as a function of radius and a radially constant relative 
fraction of ${\rm SN\;Ia}$, $f=0.15$. This fraction was chosen to lie within
the range suggested by the Si/Fe abundance ratio profile and it only affects the normalization of the predicted profiles. 
The explanation of the observed range of values in the radial profiles of the Si/Fe and S/Fe ratios by a radial trend in the IMF would require extremely steep IMF at larger radii. 
Moreover, this scenario would produce very large central increases of the Ne/Fe and Mg/Fe ratios, which we would clearly detect even given the systematic uncertainties in the Ne and Mg abundance determinations. 

While we cannot rule out a small radial trend in the stellar IMF, it cannot be responsible for the large observed ranges in the abundance ratios. 

\subsubsection{Radial trends in the ${\rm SN_{CC}}$ pre-enrichment}
\label{section:enrich}

\begin{figure}
\scalebox{0.43}{\includegraphics[angle=270]{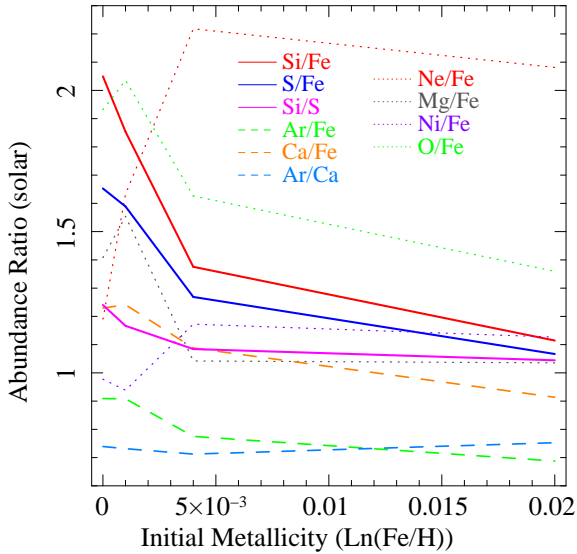}}
\caption{Predicted abundance ratios as a function of the initial
metallicity of ${\rm SN_{CC}}$ progenitors. The absolute metallicity of $Z_i=0.02$ represents the Solar metallicity. Based on ${\rm SN_{CC}}$ yields by Nomoto \etal (2006) and assuming that SN~Ia (WDD2 model by Iwamoto et al. 1999) make up 15 per cent of all supernovae.
}
\label{fig:metals}
\end{figure}

We examine the effects of a possible radial trend in the ${\rm SN_{CC}}$ pre-enrichment, i.e. the metallicity of the stars that produce the supernovae on the radial profiles of abundance ratios in the ICM. 

Fig. \ref{fig:metals} shows the predicted abundance ratios assuming that the metallicity of the ${\rm SN_{CC}}$ progenitors is increasing as a function of radius. The yields of ${\rm SN_{CC}}$ as a function of the initial metallicity are from Nomoto \etal (2006) and the absolute metallicity of  $Z_i=0.02$ is equal to Solar. We assume a constant relative fraction of ${\rm SN\;Ia}$ ($f=0.15$) and the Salpeter IMF. 
The plot shows that ${\rm SN_{CC}}$ with a lower initial metallicity produce higher Si/Fe and S/Fe ratios.

Enrichment from infalling, low-entropy systems, which may be dominated by ${\rm SN_{CC}}$ 
with relatively low metallicity progenitors during starbursts, could have contributed to the observed central peaks in the abundance ratios.

We note that ${\rm SN_{CC}}$ with a very low initial metallicity ($Z_i<0.004$) would produce large Si/Fe, S/Fe, Mg/Fe, Ca/Fe, Ar/Fe, Si/S, and Ar/Ca ratios and small Ne/Fe and Ni/Fe ratios that do not match our
observations. Therefore, bulk of the metals observed in the ICM could not have been produced by extremely low metallicity stars.

\subsubsection{Diversity in the ${\rm SN\;Ia}$ population}
\label{section:diversity}

\begin{figure}
\scalebox{0.43}{\includegraphics[angle=270]{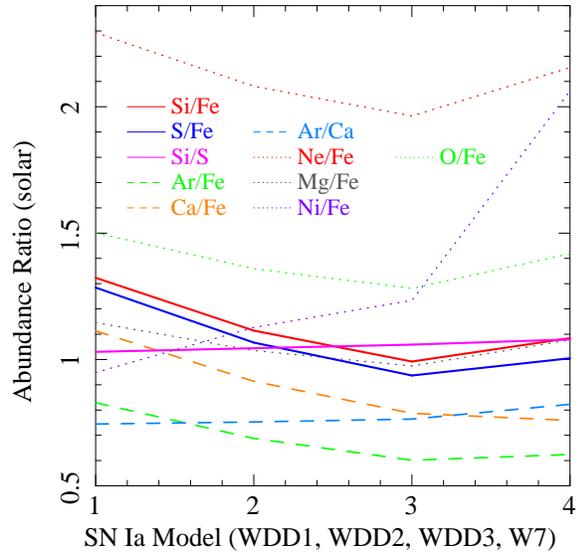}}
\caption{Predicted abundance ratios for a variety of ${\rm SN\;Ia}$ yield models. 
We considered the WDD1, WDD2, WDD3, and W7 models of Iwamoto \etal (1999).
For ${\rm SN_{CC}}$ we use the yields by Nomoto \etal (2006) and we assume that SN~Ia make up 15 per cent of all supernovae.
}
\label{fig:div}
\end{figure}

There is a growing evidence for a diversity in ${\rm SN\;Ia}$ explosions (see \eg 
Sullivan \etal 2006; Pritchet \etal 2008; Mannucci \etal 2006).
While a population of brighter SN~Ia with a slow luminosity decline is more common in late-type spiral and irregular galaxies with recent star formation
(indicating a short delay time between their formation and the explosion), a fainter and more rapidly decaying population of SN Ia is more common in early-type galaxies (Hamuy \etal 1996; Ivanov \etal 2000). This diversity should be reflected in the abundance yields with the brighter SN~Ia producing more Ni and less Si-group elements than the fainter ones.

Based on early work with {\it XMM-Newton}, Finoguenov \etal (2002) argue that the diversity in the SN~Ia population would explain the distribution of chemical elements
in the Virgo Cluster.

The variation of the peak brightness, which correlates with the production of $^{56}$Ni and anti-correlates with the production of Si-group elements, can also
be explained in the framework of the delayed detonation models by a variation of the deflagration-to-detonation transition density (transition from subsonic to supersonic
flame velocities).
Fig. \ref{fig:div} presents the expected yields for a variety of ${\rm SN\;Ia}$ explosion models from Iwamoto \etal (1999), with the WDD1, WDD2, WDD3 and W7 models on the x-axis. 
We assume a constant relative fraction of ${\rm SN\;Ia}$, $f=0.15$. 
The W7 model represents a pure deflagration explosion mechanism. The WDD models represent delayed-detonation explosions and the last digit indicates 
the density at which the flame velocity becomes supersonic (deflagration-to-detonation transition density) in units of $10^7$~g~cm$^{-3}$.
This transition density is likely dependent on the composition of the progenitor (see Jackson \etal 2010). 

Matching our observed Si/Fe and S/Fe profiles, given the existing ${\rm SN\;Ia}$ yields, would require that the centre of the galaxy has been enriched almost solely by WDD1 supernovae while the outer regions almost solely by WDD3. The contribution by SN~Ia with longer delay times and larger Si/Fe ratio would be the largest in the centre of the galaxy, but  any realistic enrichment scenario predicts an enrichment by a mixture of different types of SN~Ia at all radii. 

This model predicts a large central increase in the Ar/Fe and Ca/Fe abundance ratios that seem to be in conflict with the observed relatively flat profiles. The predicted Mg/Fe profile is relatively flat in agreement with observations. 
The strongest prediction of this model is the $\sim30$ per cent rise in the Ni/Fe abundance ratio with the increasing radius. Our observed Ni/Fe profile, which is unfortunately dominated by systematic uncertainties, suggests a relatively flat radial distribution.

\subsection{The O abundances}

The O/Fe ratio of $0.60\pm0.03$ Solar determined in the centre of M~87 with the {\it XMM-Newton} Reflection Grating Spectrometers (Werner et al. 2006b; these data resolve the \ion{O}{viii} line and the individual lines of the Fe-L complex)  is significantly lower than the values predicted by the proposed enrichment scenarios. 
These best fit O/Fe ratios are consistent with the values determined using the CCD type detectors on  {\it XMM-Newton} (Matsushita et al. 2003).
Either the measurements strongly underestimate the O abundance or some key aspect of the chemical enrichment of the hot ICM/ISM is not understood. 
Using the recently updated AtomDB atomic database, the best fitting O/Fe ratios are 50 per cent larger
compared to the previous version, indicating that at least part of the discrepancy might be a modeling 
issue.
Furthermore, all of our proposed enrichment scenarios are incompatible with the rising O/Fe abundance profile reported from {\it XMM-Newton} (B\"ohringer et al. 2001, Finoguenov et al. 2002, Matsushita et al. 2003). However, measurements of the \ion{O}{viii} line emission at $E\sim0.65$~keV with CCD detectors suffer from significant systematic uncertainties due to a combination of limited spectral resolution, residual gain uncertainties, coupled with incomplete modelling of the detector oxygen edge and possible incomplete subtraction of the \ion{O}{viii} line emission from the Galactic foreground (which could bias the O abundance measurement in the outskirts of M~87 high). These systematic uncertainties force us to treat all current O abundance measurements with caution. More robust measurements of the O/Fe profile will be possible with the calorimeters on the {\it Astro-H} satellite.

\section{Summary}
Using a deep (574~ks) {\it Chandra} observation of M\,87, we performed the best measurements to date of the radial distributions of metals in the ambient central ICM of the Virgo Cluster. We conclude that:
\begin{itemize}

\item
The abundance profiles of Fe, Si, S, Ar, Ca, Ne, Mg, and Ni are all centrally peaked. 

\item
The abundance profiles of Si and S are more centrally peaked than Fe, challenging the standard picture of chemical enrichment in galaxy clusters, wherein SN~Ia products are thought to dominate the central enrichment. Rather, despite a negligible current star formation rate in M\,87 and the continuing enrichment by SN~Ia, the integrated relative contribution of core collapse supernovae to the enrichment is higher in the central low entropy core than in the surrounding ICM. 
The observed abundance patterns are most likely due to one or more of the following processes: continuing enrichment by winds of a stellar population which has been pre-enriched mainly by SN$_{\rm CC}$ products; intermittent formation of massive stars in the central cooling core on time scales shorter than the continuing enrichment by SN~Ia; or strong early SN$_{\rm CC}$ enrichment of the lowest entropy X-ray emitting gas that is subsequently not well mixed and does not cool out of the hot ICM phase.

\item 
Other processes, such as  stellar initial mass function that changes with radius, changes in the pre-enrichment of core-collapse supernova progenitors, and diversity in the elemental yields of SN~Ia, might have also contributed to the observed radial profiles.  

\item
Although systematic uncertainties prevent us from measuring the O abundance robustly, indications are that it is about 2 times lower than predicted by the enrichment models.

\end{itemize}

\section{Acknowledgments}

We thank R.G. Morris for computational support.
We thank J.A. Irwin and J. de Plaa for stimulating discussions. We thank the anonymous referee for the important suggestions which significantly improved the paper. 
N. Werner and A. Simionescu
were supported by the National Aeronautics and Space Administration
through Chandra/Einstein Postdoctoral Fellowship Award Number PF8-90056 and
PF9-00070 issued by the Chandra X-ray Observatory Center, which is operated
by the Smithsonian Astrophysical Observatory for and on behalf of the National
Aeronautics and Space Administration under contract NAS8-03060. This work was
supported in part by the US Department of Energy under contract number
DE-AC02-76SF00515. All computational analysis was carried out using the
KIPAC XOC compute cluster at Stanford University and the Stanford
Linear Accelerator Center (SLAC).


\begin{thebibliography}{}
\bibitem{} Arnaud K.A., 1996, in Astronomical Data Analysis Software and Systems V, eds. Jacoby G. and Barnes J., ASP Conf. Series volume 101, p17
\bibitem{} Badenes, C., Bravo, E., Borkowski, K. J., Dominguez, I., 2003 ApJ, 593, 358
\bibitem{} Badenes C., Borkowski K.J., Hughes J.P., Hwang U., Bravo E., 2006, ApJ, 645, 1373
\bibitem{} Balucinska-Church M., McCammon D., 1992, ApJ, 400, 699
\bibitem{} B\"ohringer H. \etal, 2001, A\&A, 365L, 181
\bibitem{} B\"ohriner H., Matsushita K., Churazov E., Finoguenov A., Ikebe Y.,
   2004, A\&A, 416L, 21
\bibitem{} B\"ohringer H., Werner N., 2010, A\&ARv, 18, 127
\bibitem{} Ciotti L., D'ercole A., Pellegrini S., Renzini A., 1991, ApJ, 376, 380
\bibitem{} De Grandi S., Ettori S., Longhetti M., Molendi S., 2004, A\&A, 419, 7
\bibitem{} de Plaa J., Werner N., Bykov A.M., Kaastra J.S., M\'endez M., Vink
   J., Bleeker J.A.M., Bonamente M., Peterson J.R., 2006, A\&A, 452, 397
\bibitem{} de Plaa J., Werner N., Bleeker J.A.M., Vink J., Kaastra J.S., 
M\`endez M., 2007, A\&A, 465, 345
\bibitem{} Dupke R.A., White R.E., 2000, ApJ, 528, 139
\bibitem{} Durret F., Lima Neto G.B., Forman W., 2005, A\&A, 432, 809
\bibitem{} Ehlert S., et al. 2011, MNRAS, 411, 1641
\bibitem{} Finoguenov A., Matsushita K., B\"ohringer H., Ikebe Y., Arnaud M.,
   2002, A\&A, 381, 21
\bibitem{} Gallagher J.S., Garnavich P.M., Berlind P., Challis P., Jha S.,
   Kirshner R.P., 2005, ApJ, 634, 210
\bibitem{} Gavazzi G., Donati A., Cucciati O., Sabatini S., Boselli A., Davies J.,
   Zibetti S., 2005, A\&A, 430, 411
\bibitem{} Hamuy M., Phillips M.M., Suntzeff N.B., Schommer R.A., Maza J., 
   Smith R.C., Lira P., Aviles R., 1996, AJ, 112, 2438
\bibitem{} Ivanov V.D., Hamuy M., Pinto P.A., 2000, ApJ, 542, 588
\bibitem{} Iwamoto K., Brachwitz F., Nomoto K., Kishimoto N., Umeda H., Hix
   W.R., Thielemann F., 1999, ApJS, 125, 439
\bibitem{} Jackson A.P., Calder A.C., Townsley D.M., Chamulak D.A., Brown E.F.,
   Timmes F.X., 2010, ApJ, 720, 99
\bibitem{} Kaastra J.S., Mewe R., 1993, Legacy, 3, 16
\bibitem{} Kalberla P.M., Burton W.B., Hartmann Dap, Arnal E.M., Bajaja E.,
   Morras R., Poeppel W.G.L., 2005, A\&A, 440, 775
\bibitem{} Kobayashi, C., \& Arimoto, N. 1999, ApJ, 527, 573
\bibitem{} Liedahl D.A., Osterheld A.L., Goldstein W.H., 1995, ApJ, 438L, 115
\bibitem{} Lodders K., 2003, ApJ, 591, 1220
\bibitem{} Mannucci F., Della Valle M., Panagia N., 2006, MNRAS, 370, 773
\bibitem{} Matsushita K., Finoguenov A., B\"ohringer H., 2003, A\&A, 401, 443
\bibitem{} Matsushita K., B\"ohringer H., Takahashi I., Ikebe Y., 2007, A\&A,
   462, 953
\bibitem{} McNamara, B. R., Rafferty, D. A., Birzan, L., Steiner, J., Wise, M. W., Nulsen, P. E. J., Carilli, C. L., Ryan, R., Sharma, M. 2006, ApJ, 648, 164
\bibitem{} Million E.T., Werner N., Simionescu A., Allen S.W., Nulsen P.E.J.,
   Fabian A.C., B\"ohringer H., Sanders J.S., 2010, MNRAS, 407, 2046
\bibitem{} Mitchell R.J., Culhane J.L., Davison P.J.N., Ives J.C., 1976, MNRAS,
   175, 29
\bibitem{} Nomoto K., Tominaga N., Umeda H., Kobayashi C., Maeda K., 2006, 
   NuPhA, 777, 424
\bibitem{} Pritchet C.J., Howell D.A., Sullivan M., 2008, ApJ, 683L, 25
\bibitem{} Ogrean G. A., Hatch N. A., Simionescu A., B\"ohringer H., Br\"uggen M., Fabian A. C., Werner N., 2010, MNRAS, 406, 354
\bibitem{} Sanders J.S., Fabian A.C., Allen S.W., Schmidt R.W., 2004, MNRAS,
   349, 952
\bibitem{} Sanders J.S., Fabian A.C., 2006, MNRAS, 371, 1483
\bibitem{} Sato K., Tokoi K., Matsushita K., Ishisaki Y., Yamasaki N.Y.,
   Ishida M., Ohashi T., 2007, ApJ, 667L, 41
\bibitem{} Sato K., Matsushita K., Ishisaki Y., Yamasaki N.Y., Ishida M., 
   Sasaki S., Ohashi T., 2008, PASJ, 60, 333
\bibitem{} Serlemitsos P.J., Smith B.W., Boldt E.A., Holt S.S., Swank J.H.,
   1977, ApJ, 211L, 63
\bibitem{} Simionescu A., Werner N., B\"ohringer H., Kaastra J.S., Finoguenov A.,  
   Br\"uggen M., Nulsen P.E.J., 2009, A\&A, 493, 409
\bibitem{} Simionescu A., Werner N., Forman W.R., Miller E.D., Takei Y.,
   B\"ohringer H., Churazov E., Nulsen P.E.J., 2010, MNRAS, 405, 91
\bibitem{} Smith R.K., Brickhouse N.S., Liedahl D.A., Raymond J.C., 2001, ApJ,
   556L, 91
\bibitem{} Sullivan M. \etal 2006, ApJ, 648, 868
\bibitem{} Tamura T., Bleeker J.A.M., Kaastra J.S., Ferrigno C., Molendi S.,
   2001, A\&A, 379, 107
\bibitem{} Tamura T., Kaastra J.S., den Herder J.W.A., Bleeker J.A.M., 
   Peterson J.R., 2004, A\&A, 420, 135
\bibitem{} Tonry J.L., Dressler A., Bakeslee J.P., Ajhar E.A., Fletcher A.B., 
   Luppino G.A., Metzger M.R., Moore C.B., 2001, ApJ, 546, 681
\bibitem{} Werner N., de Plaa J., Kaastra J.S., Vink J., Bleeker J.A.M., 
   Tamura T., Peterson J.R., Verbunt F., 2006a, A\&A, 449, 475
\bibitem{} Werner N., B\"ohringer H., Kaastra J.S., de Plaa J., Simionescu A.,
  Vink J., 2006b, A\&A, 459, 353
\bibitem{} Werner N., Durret F., Ohashi T., Schinder S., Wiersma R.P.C., 2008,
   SSRv, 134, 337
\bibitem{} Werner N. \etal, 2010, MNRAS, 407, 2063
\end{thebibliography}
\end{document}